\documentclass[journal,11pt,onecolumn,draftclsnofoot,]{IEEEtran}
\usepackage{amsfonts}
\usepackage{amsmath}
\usepackage{amssymb}
\usepackage{graphicx}
\usepackage{cite}
\usepackage{algorithmic}
\usepackage{algorithm}
\usepackage{array}%
\usepackage{hyperref}
\usepackage{amsmath}
\usepackage{enumerate}
\usepackage{xcolor}
\usepackage{color}

\ifCLASSINFOpdf
\else
\fi

\newtheorem{lm}{Lemma}
\RequirePackage[singlelinecheck=off]{caption}
\newcommand{\rn}[1]{%
  \textup{\uppercase\expandafter{\romannumeral#1}}%
}
\newtheorem{defn}{Definition}

\begin{document}
%
\title{{Privacy-guaranteed Two-Agent Interactions Using Information-Theoretic Mechanisms}}
%
%
%

\author{Bahman~Moraffah,~\IEEEmembership{Student,~IEEE,}
        and~Lalitha~Sankar,~\IEEEmembership{Senior Member,~IEEE}
\thanks{B. Moraffah and L. Sankar are with the Department
of Electrical, Computer, and Energy Engineering, Arizona State University, Tempe,
AZ, 85287 USA e-mail: {\{bahman.moraffah,lalithasankar\}}@asu.edu. This work was supported in part by the National Science Foundation CAREER award CCF-1350914. A part of this paper was presented at the $53^{rd}$ Allerton conference on Communication, Control and Computing, Monticello, IL, Sep. 29-Oct. 2, 2015. }}

\maketitle

\begin{abstract}
This paper introduces a multi-round interaction problem with privacy constraints between two agents 
that observe correlated data. The data is assumed to have both public and private features and the goal of the 
interaction is to share the public data subject to utility constraints (bounds on distortion of public feature) while
ensuring bounds on the information leakage of the private data at the other agent. The agents alternately share data with 
one another for a total of $K$ rounds such that each agent initiates sharing over $K/2$ rounds. The interactions are modeled as a collection 
of $K$ random mechanisms (mappings), one for each round. The goal is to jointly design the $K$ private
mechanisms to determine the set of all achievable distortion-leakage pairs at each agent. 
Arguing that an mutual information-based leakage metric can be 
appropriate for streaming data settings, this paper: (i) determines the set of all achievable distortion-leakage tuples ; (ii) shows that the $K$
mechanisms allow for precisely composing the total privacy budget over $K$ rounds without loss; and (ii) develops conditions under which 
interaction reduces the net leakage at both agents and illustrates it for a specific class of sources. The paper then 
focuses on log-loss distortion to better understand the effect on leakage of using a commonly 
used utility metric in learning theory. The resulting interaction problem leads 
to a non-convex sum-leakage-distortion optimization problem that can be viewed as an interactive version of the information bottleneck problem. A new \textit{merge-and-search} 
algorithm that extends the classical agglomerative information bottleneck algorithm to the interactive setting is introduced to determine a provable locally optimal solution. Finally, the benefit of interaction under log-loss is illustrated for 
specific source classes and the optimality of one-shot is proved for Gaussian sources under both mean-square and log-loss distortions constraints.

\end{abstract}

\begin{IEEEkeywords}
Private interactive mechanism, distortion-leakage tradeoff, log-loss distortion, composition rule.
\end{IEEEkeywords}

%
\IEEEpeerreviewmaketitle

\section{Introduction}
\label{sec:sec1}
\IEEEPARstart{C}{onsider} an electric power system in which systems operators that manage specific sub-areas of the network share measurements with each other to obtain precise estimates of the underlying system state, i.e., complex voltages. Despite the need for such sharing and the value of high fidelity state estimates, such sharing is often limited due to privacy considerations; in the process of sharing measurements the operators do not wish to leak information about a subset of their internal states. However, since the measurements need to be shared, and often multiple times due to the iterative nature of power systems state estimation, it is crucial to understand: (a) the effect of applying privacy-preserving mechanisms on both the utility of estimation and leakage of the private data; and (b) the effect of multiple rounds of interaction and sharing on the net leakage.

Privacy in such a distributed ``competitive" context is different from the traditional statistical database privacy setting in which data is published to ensure statistical value while ensuring that the privacy of any individual in the database is not comprised. In this database context, differential privacy with guarantees on the worst-case privacy leakage has emerged as a strong formalism \cite{Dwork2006}. However, in many data sharing settings, such as the above-mentioned electric power system example as well as other streaming data settings (e.g., sensors networks, IoT, even electronic medical records, etc), the data stream as a whole has private and public features that need to be hidden and revealed, respectively. {In such settings where the privacy threat is primarily about inference, a statistical approach using an information-theoretic privacy framework can capture the correlation between the public and private features. In fact, the current definitions of differential privacy which focus on protecting individual privacy cannot be applied easily to study the tradeoffs for multi-feature privacy vs. inference problems.}

To this end, we consider a two-way interactive data sharing setting with two agents. Each agent generates an $n$-length independent and identically distributed (i.i.d.) sequence of public and private data; data at the two agents are assumed to be correlated as is generally the case in such distributed settings. Each agent wishes to share a function of its public data with the other agent to satisfy a desired measure of utility (e.g., via a distortion function) while ensuring that a mutual information based leakage of its private data is constrained over $K$ rounds of communications.

Formally, an information-theoretic privacy mechanism is a randomizing function that maps the public data from a data source to an output (\textit{revealed/released} data); any such mapping will achieve a certain utility, quantified via a desired distortion function, and leakage of private data quantified via average mutual information. In the interactive setting, we allow for a total of $K$ rounds of data sharing ($K/2$ rounds per agent) and introduce a private interactive mechanism as a collection of $K$ random mappings. From both a theoretical and an application viewpoint, it is of much interest to understand whether interaction reduces privacy leakage or if a single round of data sharing suffices for a fixed privacy budget (leakage constraint). 
\subsection{Related Work}
An information-theoretic formulation of the utility-privacy tradeoff problem was introduced in \cite{Sankar2013} for the one-shot data publishing setting and has also been studied in \cite{PinCalmon2012, Salamatian2013}. For the interactive setting, \cite{Sankar2011} determines the largest achievable utility-privacy tradeoff region for a two-agent system for a class of Gaussian sources and mean-squared distortion functions at both agents. In contrast, the focus in this paper is on {both discrete and Gaussian memoryless sources and appropriate classes of distortion functions.} 

For a one-way non-interactive setting, in \cite{Makhdoumi2014} Makhdoumi \textit{et al.} introduce an algorithm based on the agglomerative information bottleneck algorithm to compute the risk-distortion tradeoff for logarithmic loss based privacy and distortion functions that they refer to as the privacy funnel problem. We will henceforth refer to the generalization of the privacy funnel problem for the interactive case that we study here as the \textit{interactive privacy funnel problem}. More recently, in \cite{Vera2015} Vera \textit{et al.} study the rate-relevance region for an interactive two-agent information bottleneck problem. 

It is worth noting that the problem at hand also falls under the purview of multiparty computation; in this context, recently, in \cite{Kairouz2014} Kairouz \textit{et al.} prove the optimality of one-shot interactions for binary sources using a differentially private data sharing mechanism. Our information-theoretic approach considers general sources and distortions as well as public and private data for two agents and shows that in general a data source can leak less over multiple rounds for a fixed distortion. Furthermore, while secure multiparty computation (SMC) is often considered a recourse to such interactive data sharing setups (see for example, \cite{Canetti2002}), the complexity of SMC implementations and the rise of many cloud-based applications with demands for real-time distributed data processing suggests the need for alternative privacy-guaranteeing approaches as motivated in \cite{Kairouz2014}. 

We also note that the interactive formulation studied here shares some similarities with interactive source coding problems introduced in \cite{Kaspi1985} and further studied in \cite{MA2011} and \cite{MA2012}. {However, unlike classical source coding, in our model each source does not `code' its data sequence but rather maps it to an intermediate  `revealed' sequence in each round such that at the end of $K$ rounds, the receiver agent `reconstructs' a typical data sequence using all the information at its disposal}. We also note that our assumption of memoryless sources leads to single-letter expressions for (mutual) information leakage as a function of the distortion pairs which bears similarities to the rate-distortion function in source coding setup. We exploit this similarity to determine the conditions under which interaction benefits leakage in a manner similar to that done for interactive source coding problem by Ma \textit{et. al.} in \cite{MA2013}. It is crucial to note that, in contrast to the traditional interactive source coding setup with rate and distortion constraints, here the leakage and distortion constraints are on different aspects of the source, namely, the private and public features, respectively. Thus, it is unclear \textit{a priori} if multiple rounds of interaction can reduce leakage or may worsen it.
\subsection{Our Contributions} {In this paper, we consider discrete memoryless correlated sources at the two agents and determine the set of all possible leakage-distortion tuples achievable at both agents over $K$ rounds of interaction (Section~\ref{sec:sec2}); for jointly Gaussian sources with quadratic distortion constraints we show the optimality of one-shot privacy mechanisms. In this same section, we also highlight how an information-theoretic approach naturally lends itself to composing privacy optimally over multiple interactions without any cumulative loss; we complete this section by illustrating the advantage of interaction for specific two-agent source and distortion models.  In Section~\ref{sec:sec3}, we determine the conditions under which interaction helps.} We then focus on a specific class of distortion functions, namely, log-loss distortion (Section~\ref{sec:sec6}), which is often used as a utility function in machine learning applications. Our motivation for this model stems from the fact that the intermediate soft decoding characteristic of many interactive systems is well captured by log-loss distortion in which each agent continually refines its belief of the data to be estimated/inferred with each interaction. We show that the resulting interactive privacy funnel problem is a dual of an interactive information bottleneck problem, and analogously, involves optimization over a non-convex probability space; to this end, we extend the agglomerative information bottleneck algorithm appropriately for the two-agent interactive case that we call the agglomerative interactive privacy algorithm. We  show that for Gaussian sources with log-loss distortion, one-shot data sharing is optimal; in contrast, we also prove that, in general, there always exists a pair of distributed correlated (non-Gaussian) sources for which interaction helps under log-loss. We illustrate our results using publicly available census data (Section~\ref{sec:sec7}) and conclude in Section~\ref{sec:sec8}.

Preliminary work on the achievable distortion-leakage region for the two-agent interaction problem with privacy constraints studied here is developed by the authors in \cite{Moraffah2015}. Furthermore, in \cite{Moraffah2015}, the authors present an example to illustrate the advantage of interactions to reduce leakage as well as study the leakage-distortion tradeoffs under log-loss distortion; to this end, \cite{Moraffah2015} introduces the interactive version of the privacy funnel problem and the related `merge-and-search' algorithm. These above-mentioned details are also covered in this paper with the prime difference being that, unlike \cite{Moraffah2015}, this paper includes the detailed proofs for all theorems and intermediate results. Additionally, this paper also develops the following with detailed proofs: (i) rigorous testable conditions under which interaction helps; (ii) a composition theorem that clarifies how an optimal mechanism composes a net leakage budget amongst $K$ rounds of interaction; and finally (iii) a detailed example to illustrate that interaction reduces leakage under log-loss distortion for a large class of binary sources.

Finally, we also briefly comment on the source and mechanism model considered here and place it context of related work. We assume that the datasets at each agent are large, i.e., the agents, if needed, could empirically evaluate the source distributions to design the interactive mechanisms. Furthermore, the data sources are assumed to be memoryless, i.e., the data of each user corresponding to a row of the dataset is independent of that of other users; however, the public and private features of each user (in any row) are correlated. We assume that the privacy mechanism over all rounds of interaction is general and not necessarily memoryless; in fact, we use the tools of asymptotic information theory to show that memoryless mechanisms suffice for memoryless sources. Many related privacy approaches implicitly assume memoryless mechanisms \cite{PinCalmon2012, Salamatian2013} thereby modeling the problem as one of ``local privacy" wherein each user applies the same mechanism independently to their own data (see for example, \cite{Duchi2013}). The implicitness comes from the fact that these works assume that the statistics of the data for every user in the dataset is known and follows the same distribution thus allowing the use of a single mechanism locally; in contrast, we show this explicitly here. 

\textit{Notation}: We use upper-case letters to denote random variable and lower-case letters to denote realizations of random variables. Superscripts are used to denote the length of a vector. {We write Var to denote the variance of a random variable and for conditional variance also use the expectation operator $\mathbb{E}$ as $\mathbb{E}[Var(\cdot|\cdot)]$.} We write $Ber(p)$ to denote the Bernoulli distribution with parameter $p$ and write $(X,Y)\sim \text{DSBS}(p)$ to denote a doubly symmetric binary source with crossover probability$P(0,1)=P(1,0)=\frac{p}{2}$. We write $H(X)$ and $I(X;Y)$ to denote the entropy and mutual information, respectively. We also interchangeably use the notation $H(P_X)$ for $X\sim P_X$. We write $D=(D_1,D_2)$ to denote the distortion vector $D$.
\vspace{-.1cm}
\section{System Model and Interactive Mechanism}
\label{sec:sec2}
\begin{figure}[htbp]
	\begin{center}
	\includegraphics[width=10cm,keepaspectratio]{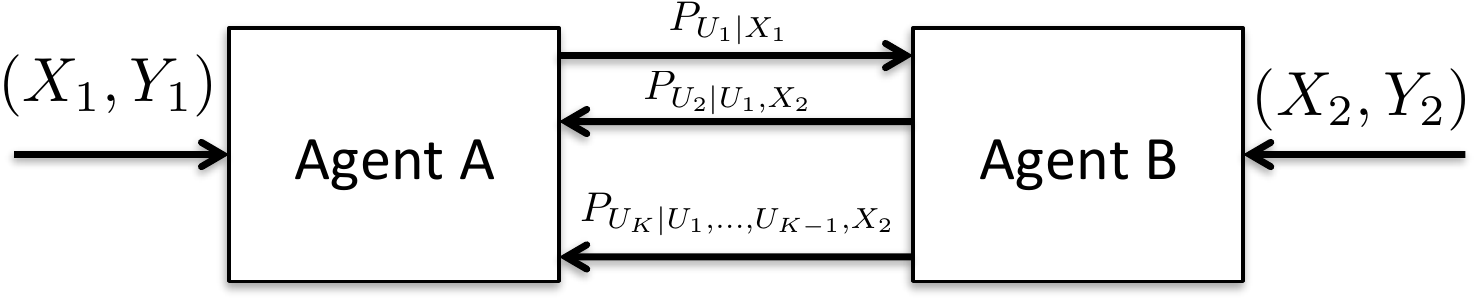}
		\caption{K-interactive Privacy Model.}
		\label{fig:fig1}
	\end{center}\vspace{-0.5cm}
\end{figure}

{Our problem consists of a discrete source (e.g., the electric power system) that generates $n$-length i.i.d. sequences $(X_1^n,Y_1^n,X_2^n,Y_2^n) \in (\mathcal{X}_1^n,\mathcal{Y}_1^n,\mathcal{X}_2^n,\mathcal{Y}_2^n)$ with $(X_{1i},Y_{1i}, X_{2i},Y_{2i})\sim P_{X_{1},Y_{1}, X_{2},Y_{2}}$, for all $i=1,2,...,n$. These sequences are partially observed at two agents that interact with one another as shown in Fig.~\ref{fig:fig1} such that the two agents $A$ and $B$ observe $n$-length sequences $(X_1^n,Y_1^n)$ and $(X_2^n,Y_2^n)$, respectively.} The public data at both agents are denoted by $X^n_{(\cdot)}$ and the correlated private data by $Y^n_{(\cdot)}$. Furthermore, we assume that the private data is hidden and can only be leaked through the public data. 
We consider a $K$-round interactive protocol in which, without loss of generality, we assume that agent A initiates the interaction and $K$ is even. A $K$-interactive privacy mechanism is given by $(n,K,\{P_{1i}\}_{i=1}^{K/2},\{P_{2i}\}_{i=1}^{K/2}, D_1, D_2, L_1, L_2)$ as a collection of  $K$ probabilistic mappings such that agent A shares data in the odd rounds beginning with round 1 and agent B shares in the even rounds.  A privacy mechanism $P_{1i}$ for agent A used in  the $(2i-1)$-th round, $i\in\{1,2,\dots,\frac{K}{2}\}$, is a mapping from its public data  sequence and all prior sequences revealed from agent B. Thus, in round 1, $P_{11} :\mathcal{X}^n_1\to\mathcal{U}^n_1$, where $\mathcal{U}^n_1$ is the revealed set when a sequence$\ U^n_1$ is shared via $P_{11}$. For the odd rounds $i=3,\dots,K-1$, the mechanism used by agent A is
\begin{equation}
\label{eq:odd_mech}
 P_{1,\frac{i+1}{2}} : (\mathcal{X}^n_1,\mathcal{U}^n_1,\mathcal{U}^n_2,\dots,\mathcal{U}^n_{i-1})\to \mathcal{U}^n_i.
 \end{equation}
 Similarly, agent B in even rounds $i$, $i\in\{2,4,\dots,K\}$ uses its public data and the prior data sequences revealed from agent A and maps them via a privacy mechanism 
 \begin{equation}
\label{eq:even_mech}
 P_{2,\frac{i}{2}} : (\mathcal{X}^n_2,\mathcal{U}^n_1, \dots,\mathcal{U}^n_{i-1})\to \mathcal{U}^n_i.
 \end{equation}

{From (\ref{eq:odd_mech}) and (\ref{eq:even_mech}), we see that our model assumes that the private sequences $Y^n_1$ and $Y^n_2$ are not explicitly involved in the mapping such that in the $(2k-1)$-{th} and $2k$-{th} rounds, $k=1,2,...,K/2$, respectively, $Y^n_{2}\leftrightarrow (U^n_{1},\dots, U^n_{2k-2}, X^n_{1})\leftrightarrow U^n_{2k-1}$ and $Y^n_{1}\leftrightarrow (U^n_{1},\dots, U^n_{2k-1}, X^n_{2})\leftrightarrow U^n_{2k}$ form Markov chains.  This is because any dependence of agent A's $U^n_k$'s on $Y_2^n$ is captured via the $U^n_i$'s from agent B and vice versa. Thus, in any round, conditioned on all data that an agent has until then, what the agent transmits is independent of the private data at the other agent. Our assumption is motivated by the fact that private data is not, in general, accessible and models inferred features that are not known \textit{a priori}. Our model is motivated by the example we have alluded to earlier, that of operators sharing measurement data in the electric power grid which could lead to estimation of each other's (private) system state; more broadly, our model captures any interactive application in which personal habits or preferences not known or observed directly $Y^n_{(\cdot)}$ can only be inferred from the data collected ($X^n_{(\cdot)}$) or shared ($\hat{X}^n_{(\cdot)}$). Furthermore, since our problem model involves a sequence of $K$ random mappings, the model directly includes $K$ auxiliary random variables $U_1,\ldots,U_K$, such that their $n$-length sequences are the outputs of the privacy mechanism over $K$ rounds. While these auxiliary variables are required to model the problem, it is unclear \textit{a priori} what the cardinalities of their support set should be, and thus, one needs to develop bounds on them in the process of determining the largest achievable utility-privacy tradeoff region. To this end, we use classical information-theoretic methods to obtain bounds on the cardinalities. It is worth noting that bounds on the cardinalities of the outputs of privacy mechanisms are also seen in problems involving other privacy mechanisms such as differential privacy (e.g., \cite{Kairouz2014a}).}

At the end of $K$ rounds, agents A and B reconstruct sequences $\hat{X}^n_{2} = g_2({X^n_{1}}, U^n_1,\dots, U^n_K)$ and $\hat{X}^n_{1} = g_1({X^n_{2}}, U^n_1,\dots, U^n_K)$, respectively, where $g_1$ and $g_2$ are appropriately chosen reconstruction functions. The set of mechanism pairs  $\{P_{1j}, P_{2j}\}_{j=1}^{\frac{K}{2}}$ is chosen to satisfy

\begin{subequations} 
{{
 \begin{align}
     \limsup\limits_{n\rightarrow\infty}\frac{1}{n}&\sum_{i=1}^{n}\mathbb{E}(d_1(X_{1i},\hat{X}_{1i})) \leq D_1\label{achievable requirement1} \\
      \limsup\limits_{n\rightarrow\infty}  \frac{1}{n}&\sum_{i=1}^{n}\mathbb{E}(d_2(X_{2i},\hat{X}_{2i}))\leq D_2\label{achievable requirement2}\\
       \limsup\limits_{n\rightarrow\infty} \frac{1}{n}&I(Y^n_1; U^n_1,\dots, U^n_K, X^n_2)\leq L_1\label{achievable requirement3}\\
        \limsup\limits_{n\rightarrow\infty}\frac{1}{n}&I(Y^n_2; U^n_1,\dots, U^n_K, X^n_1)\leq L_2\label{achievable requirement4}
       \end{align}}}
       \end{subequations}
where $d_1(\cdot,\cdot)$ and $d_2(\cdot,\cdot)$ are the given distortion measures.

The utility-privacy tradeoff region is the set of all $(L_1,D_1,L_2,D_2)$ tuples for which a privacy mechanism exists and is given by the following theorem.

\newtheorem{thm}{Theorem}
\begin{thm} \label{m:main}
For a target distortion pair $(D_1, D_2)$ and a $K$-round interactive privacy mechanism, the utility-privacy tradeoff region is the set of all $(L_1,L_2,D_1,D_2)$ tuples that satisfy
\begin{flalign}
\ & L_1\geq L_{U,1}(D_1,D_2) = I(Y_1; U_1,\dots, U_K, X_2),\notag\label{eq:LU_def} \\
\ & L_2\geq L_{U,2}(D_1,D_2) = I(Y_2; U_1,\dots, U_K, X_1),\notag \\
\ &{{\mathbb{E}}}(d_1(X_1,\hat{X}_1))\leq D_1,\notag\\
\ &{{\mathbb{E}}}(d_2(X_2,\hat{X}_2))\leq D_2\
\end{flalign}
such that for all $k$, the following Markov chains hold:
\begin{flalign}
Y_1\leftrightarrow (U_1,\dots, U_{2k-1}, X_2)\leftrightarrow U_{2k}\label{eq:MC_1}\\
Y_2\leftrightarrow (U_1,\dots, U_{2k-2}, X_1)\leftrightarrow U_{2k-1}\label{eq:MC_2}\vspace{-0.4cm}
\end{flalign}
with $|\mathcal{U}_l|\leq|\mathcal{X}_{i_l}|.(\prod_{j=1}^{l-1}|\mathcal{U}_j|)+1$ where $i_l = 1$ if $l$ is odd and $i_l = 2$ if $l$ is even.
\label{thm: thm1}
\end{thm}
\begin{IEEEproof}
The proof details are in Appendix A. We briefly review the steps. { Achievability follows from using an i.i.d. mechanism in each round and using strong typicality (defined precisely in Appendix A) to bound the achievable leakage at both agents. The converse, on the other hand, considers a mechanism that achieves (\ref{achievable requirement1})-(\ref{achievable requirement4}) and exploits the i.i.d. nature of correlated sources to obtain single letter bounds. We also note that the Markov chains in (\ref{eq:MC_1}) and (\ref{eq:MC_2}) directly capture the fact that at each transmitting agent the data shared in the next round is independent of the private data at the receiving agent conditioned on the data available at the transmitting agent (including the data from the previous rounds). This Markovity is a result of the in turn is due to the fact that the private data is not directly used in the random mapping in each round and $Y_i^n \leftrightarrow X_i^n \leftrightarrow \hat{X}_i^n, i=1,2$.}
\end{IEEEproof}
\newtheorem{pa}{Remark}
{
\begin{pa}
Note that Theorem~\ref{thm: thm1} holds even if agent B initiates the interaction; however now $U_1$ will be the output of agent B in round 1 and $U_2$ will be the output of agent A in round 2, and so on, such that the Markov conditions are appropriate  in Theorem~\ref{thm: thm1}.  
\end{pa}}
\newtheorem{ppp}{Corollary}
\begin{ppp}
For the special case, $Y_i = X_i$, $i=1,2$, i.e., when the public and private data are the same, the region in Theorem~\ref{thm: thm1} yields the set of all achievable sum-rate and distortion $(R_1,R_2,D_1,D_2)$ tuples for the interactive source coding problem in \cite{Kaspi1985} with $R_i = L_i -I(X_1;X_2)$.
\label{ppp: coro1}\
\end{ppp}
\begin{IEEEproof}
For $X_i=Y_i$, the leakage term $L_1$ in Theorem~\ref{thm: thm1} can be written as $L_1 = I(X_1;X_2) + I(X_1;U_1,\ldots,U_K|X_2)$ such that, for the equivalent interactive source coding problem in \cite{Kaspi1985}, the source coding sum-rate $R_1$ is simply the excess information that needs to be shared beyond what can be inferred at the receiver via $I(X_1;X_2)$, i.e., $R_1 = L_1-I(X_1;X_2)$, and similarly, $R_2 = L_2-I(X_1;X_2)$. 

\end{IEEEproof}

\begin{pa}
Note that a \textit{one-shot} setting is one in which both agents share data independently and simultaneously with each other only once.
\end{pa}

Without loss of generality we assume we initiate interaction from agent $A$  such that the last round of interaction is from agent $B$ to agent $A$. We define a compact subset of a finite Euclidean space as
 \begin{flalign}
 \label{prob:probspace}
 \mathcal{P}^A_K := & \{P_{U^K|X_1,Y_1,X_2,Y_2}: P_{U^K|X_1,Y_1,X_2,Y_2} = P_{U_1|X_1}P_{U_2|U_1,X_2}\dots, P_{U_K|U^{K-1},X_2},\notag\\
 & {{\mathbb{E}}}(d_1(X_1,\hat{X}_1))\leq D_1, {{\mathbb{E}}}(d_{2}(X_2,\hat{X}_2))\leq D_2\}
 \end{flalign}

In addition to the tradeoff region, one can also focus on the net leakage over $K$ rounds.  From Theorem \ref{m:main}, the sum leakage-distortion function over $K$ rounds initiated from agent A is 
\begin{flalign}
\label{eq:ksum}
L^A_{\text{sum},K}(D_1, D_2) = \min_{P_{U^K|X_1,Y_1,X_2,Y_2}\in  \mathcal{P}^A_K } &\{I(Y_1; U_1,\dots, U_K, X_2)+I(Y_2; U_1,\dots, U_K, X_1)\}.
\end{flalign}
For the region given by Theorem~\ref{m:main} with target distortions $D_1$ and $D_2$, one can define a sum leakage over any $k$ rounds, $k = 1,2,\dots,K$. Assuming agent A initiates the interactions, we have 
\begin{equation}
L^A_{\text{sum},k}(D_1, D_2)  = \sum_{\substack {i,j=1\\ i\neq j}}^{2}I(Y_i;X_j)+ \min_{P_{U^K|X_1,Y_1,X_2,Y_2}\in  \mathcal{P}^A_K }{(\sum_{i=1}^{k}{ I(Y_1;U_i |X_2, U^{i-1})}+\sum_{i=1}^{k}{ I(Y_2;U_i |X_1, U^{i-1})})}.
\end{equation}
One can similarly define $L^B_{sum,k}$ for sum leakage over $k$ rounds originating from agent B. 
\begin{lm}
\label{lm:lm0}
 For all $k$,
\begin{equation}
\begin{aligned}
\label{eq:AA}
 L^A_{\text{sum},(k-1)}\geq L^A_{\text{sum},k}, \mbox{and}, L^B_{\text{sum},(k-1)}\geq L^B_{\text{sum},k}
 \end{aligned}
 \end{equation}
 \begin{equation}
 \label{eq:AB}
L^B_{\text{sum},(k-1)}\geq L^A_{\text{sum},k}, \mbox{and}, L^A_{\text{sum},(k-1)}\geq L^B_{\text{sum},k}. 
 \end{equation}
 
\end{lm}
\begin{IEEEproof}
The bounds in (\ref{eq:AA}) for all $k$ follow from the fact that any $(k-1)$-round interactive mechanism starting at one of the agent (e.g., A) can be considered as special case of $k$-round interactive mechanism starting at the same agent with {{$P_{U_k=0|U^{k-1},X_{(\cdot)}} = 1$ for all $(U^{k-1},X_{(\cdot)})$ sequences, i.e., a deterministic $U_k$ sequence (w.l.o.g., with entries $U_{k,i} = 0$) is sent thereby conveying no information}. The bounds in (\ref{eq:AB}) follow from the fact that any $(k-1)$-round interactive mechanism initiated at B (respectively A) can be considered as a special case of a $k$-round interactive mechanism initiated at agent A (respectively B) with {{$P_{U_1=0|X_1} = 1$ (respectively $P_{U_1=0|X_2}=1$)}}.}
\end{IEEEproof}
\begin{defn}
 $L_{sum,\infty} := \lim_{k\to\infty}L^A_{\text{sum}, k} = \lim_{k\to\infty}L^B_{\text{sum}, k}$.
\end{defn}
From the inequality (\ref{eq:AA}) in Lemma \ref{lm:lm0}, $L^A_{\text{sum}, k}$ and $L^B_{\text{sum}, k}$ are both non-increasing in $k$ and bounded from below, and thus their limits exist. Furthermore, from the inequality (\ref{eq:AB}) in Lemma \ref{lm:lm0}, $L^A_{\text{sum}, k-1}\geq L^B_{\text{sum}, k}\geq L^A_{\text{sum}, k+1}$. Thus, taking limits, since both $L^A_{\text{sum}, k}$ and $L^B_{\text{sum}, k}$ converge we have that $L_{\text{sum},\infty} := \lim_{k\to\infty}L^A_{\text{sum}, k} = \lim_{k\to\infty}L^B_{\text{sum}, k}$ and thus, we can define and compute $L_{\text{sum},\infty}$.

{\subsection{Gaussian Sources: Interactive Mechanism}}
 \label{lb:2B}
We now consider the case where the data pairs at each agent  are drawn according to bivariate Gaussian distributions, i.e., $(X_1,Y_1)\sim N(0,\Sigma_{X_1,Y_1})$, $(X_2,Y_2)\sim N(0,\Sigma_{X_2,Y_2})$, and $(X_1,X_2)\sim N(0,\Sigma_{X_1,X_2})$. For jointly Gaussian sources subject to mean square error distortion constraints, we prove that  one round of interaction suffices to achieve the utility-privacy tradeoff. 

\begin{thm} \label{thm:thm2+}
For the private interactive mechanism, the leakage-distortion region under mean square error distortion constraints consist of all tuples $(L_1, L_2, D_1, D_2)$ satisfying
\begin{flalign}
L_1\geq \frac{1}{2}\log(\frac{{\sigma}^2_{Y_1}}{\alpha^2 D_1+{\sigma}^2_{Y_1|X_1, X_2}})\\
L_2\geq \frac{1}{2}\log(\frac{{\sigma}^2_{Y_2}}{\beta^2 D_2+{\sigma}^2_{Y_2|X_1, X_2}})
\end{flalign}
where $\alpha = \frac{cov(X_1,Y_1)}{{\sigma}^2_{Y_1}}$ and $\beta = \frac{cov(X_2,Y_2)}{{\sigma}^2_{Y_2}}$ .
\end{thm}
\begin{IEEEproof}
If $(X_1,Y_1)$ is jointly Gaussian, we can write $Y_1 = \alpha X_1+ \beta X_2+Z_1$, where $Z_1$ is a zero mean Gaussian random variable independent of $X_1$ and $X_2$. 

Achievability is established by considering a single round Gaussian mechanism, i.e., the sequence $U^n_1$  is chosen such that the `test channel' from $U_1$ to $X_1$ yields $U_1 = X_1 + V_1$, where $V_1$ is a zero-mean Gaussian with variance $Q$ which is independent of the rest of the random variables. The variance $Q$ is chosen such that the reconstruction function  of $\hat{X}_1$, i.e., the minimum mean square estimate ( MMSE) of $X_1$ given $U_1$ and $X_2$, is $D$. 

To prove the converse, we have
\begin{flalign}
L_1+\epsilon\geq&\frac{1}{n}I(Y^n_1;U^n_1,\dots, U^n_K, X^n_2)\\
=&\frac{1}{n}[h(Y^n_1) - h(Y^n_1|U^n_1,\dots, U^n_K, X^n_2)]\label{eq:Gauss1}\\
=&\frac{1}{n}[nh(Y_1) -\sum_{i=1}^{n}h(Y_{1i}|U^n_1,\dots, U^n_K, X^n_2, Y^{i-1}_1)]\label{eq:Gauss2}\\
\geq&h(Y_1) - \frac{1}{n}\sum_{i=1}^{n}h(Y_{1i}|U^n_1,\dots, U^n_K,X^n_2)\label{eq:Gauss3}\\
\geq&h(Y_1) - \frac{1}{n}\sum_{i=1}^{n}\frac{1}{2}\log(2\pi e(\mathbb{E}\text{var}(Y_{1i}|U^n_1,\dots, U^n_K, X^n_2)))\label{eq:Gauss4}\\
\geq&h(Y_1) - \frac{1}{2}\log(2\pi e\frac{1}{n}\sum_{i=1}^{n}(\mathbb{E}\text{var}(Y_{1i}|U^n_1,\dots, U^n_K, X^n_2)))\label{eq:Gauss5}\\
\geq&h(Y_1) - \frac{1}{2}\log(2\pi e\frac{1}{n}\sum_{i=1}^{n}(\mathbb{E}\text{var}(\alpha X_{1i}+Z_{1i}|U^n_1,\dots, U^n_K, X^n_2)))\label{eq:Gauss6}\\
\geq &\frac{1}{2}\log(\frac{{\sigma}^2_{Y_1}}{\alpha^2 D_1+{\sigma}^2_{Y_1|X_1, X_2}}\label{eq:Gauss7})
\end{flalign}
where (\ref{eq:Gauss1}) follows from expanding the mutual information, (\ref{eq:Gauss2}) from using chain rule for entropy and the fact that the sources are i.i.d., (\ref{eq:Gauss3}) from the fact that conditioning does not increase entropy, (\ref{eq:Gauss4}) from the fact that the conditional differential entropy is maximized by a Gaussian distribution for a given variance, (\ref{eq:Gauss5}) from the concavity of the entropy function, (\ref{eq:Gauss6}) from the fact that $(X_1, Y_1)$ are jointly Gaussian, and thus, can be written as $Y_1 = \alpha X_1+ \beta X_2+Z_1$ where $Z_1$ is independent of $X_1$ and $X_2$. The final expression in (\ref{eq:Gauss7}) follows from the following facts: (i) {$Z_{1i}$ is independent of $X_{1i}$ and $X_{2i}$, and thus, $\mathbb{E}\text{var}(\alpha X_{1i}+Z_{1i}|U^n_1,\dots, U^n_K, X^n_2) = \mathbb{E}\text{var}(\alpha X_{1i}|U^n_1,\dots, U^n_K, X^n_2)+\mathbb{E}\text{var}(Z_{1i}|U^n_1,\dots, U^n_K, X^n_2) = \mathbb{E}\text{var}(\alpha X_{1i}|U^n_1,\dots, U^n_K, X^n_2)+\mathbb{E}\text{var}(Z_{1i}|h(X_1^n,X_2^n), X^n_2) = \mathbb{E}\text{var}(\alpha X_{1i}|U^n_1,\dots, U^n_K, X^n_2)+\mathbb{E}\text{var}(Z_{1i})$ where $h(\cdot)$ is a random function of $(X_1^n,X_2^n)$, and therefore, independent of $Z_{1i}$ for all $i$;} (ii) from the definition of the quadratic distortion function, $\hat{X}_1^n$ is the minimum mean square estimate of $X_1^n$ given $(U_1^n,\ldots,U_K^n,X_2^n)$, and thus, $\mathbb{E}\text{var}(X_{1i}|U_1^n,\ldots,U_K^n,X_2^n)=D_{1i}$, for all $i=1,\ldots,n$, where $D_{1i}$ is the distortion of the $i^{th}$ entry of $X_1^n$; we use this in conjunction with the fact that $\frac{1}{n}\sum_{i=1}^nD_i \le D$ to obtain the first term in the denominator of (\ref{eq:Gauss7}); (iii) since $Y_{1i}=\alpha X_{1i}+\beta X_{2i} + Z_{1i}$, $\text{var}(Z_{1i})=\mathbb{E}\text{var}(Y_{1i}|X_{1i},X_{2i})$, and thus, $\mathbb{E}\text{var}(Z_{1i}|X^n_2)=\mathbb{E}\text{var}(Y_{1i}|X_{1i},X_{2i})$ since the sources are memoryless (in fact, each instantiation of the source is independent and identically distributed); { and finally the numerator of \ref{eq:Gauss7} follows directly from the fact that the sources are jointly Gaussian distributed}. We can similarly prove that $L_2\geq \frac{1}{2}\log(\frac{{\sigma}^2_{Y_2}}{\beta^2 D_2+{\sigma}^2_{Y_2|X_1, X_2}})$.
\end{IEEEproof}

\begin{pa}
\label{remark:Gauss}
One can notice in the case that $Y_1\leftrightarrow X_1\leftrightarrow X_2\leftrightarrow Y_2$ is a Markov chain, we have $\mathbb{E}\text{Var}(Y_1|X_1, X_2) = \mathbb{E}\text{Var}(Y_1|X_1)$. 
\end{pa}

 \subsection{Composition Rules}
 \label{lb:2A}
When guaranteeing privacy, it is important to understand whether a given total leakage budget can be allocated optimally over multiple rounds such that the sum of the leakages in each round does not exceed this total. Thus, we seek to understand if the net leakage constraint can be ``composed" (or alternately decomposed) appropriately over multiple rounds. The following theorem summarizes our results. 

 \begin{thm}
For a $K$ round interactive data sharing setup between two agents A and B, the total leakage constraint $L$ can be (de-)composed into $K$ leakages, one for each round, without any loss if in each round the privacy mechanism at each agent is chosen conditioned on all data (received and known \textit{a priori}) available at each agent.
\end{thm}
\begin{IEEEproof}
We now show that the information-theoretic model presented here allows taking a total leakage budget and (de-)composing it into $K$ parts. We first observe that at the beginning of round $k$ ($k$ odd) from agent A to B, agent B has access to $(X_2,U_1,\dots,U_{k-1})$ from prior rounds and its own data. The leakage for just this $k^{th}$ round with mechanism ${P_{1,\frac{k+1}{2}}}$, for all $k=1, \dots, K/2,$ can be easily verified to be $L^{(k)}_{U,1} = I(Y_1;U_k | X_2,U_1, \dots, U_{k-1})$ (~\cite{Sankar2013}, Theorem 2). On the other hand, the net leakage at agent B over $K$ rounds is $L_{U,1}(D_1,D_2) = I(Y_1; U_1,\dots, U_K, X_2) = \sum_{i=1,i \in [1,3,\dots,K-2,K]}^{k}{ I(Y_1;U_i |X_2, U^{i-1})} = \sum_{i=1,i \in [1,3,\dots,K]}^{k}{L^{(k)}_{U,1}}$, where the even numbered terms are zero since $U_k$ for even $k$ is a mapping of $X_2$ and thus conditioning on $X_2$ provides no new information. One can similarly write the expression for leakage at agent B. 

Thus, we see that the net leakage of the private information of agent A (B) at agent B (A) is simply a sum of the leakages for each round of communication initiated at A (B) and ending at B (A), i.e., the $K$-round privacy mechanism satisfies a well desired composition property that the net leakage is not greater than the sum of the parts.  Such a composition is a direct result of the fact that the privacy mechanism in each round is chosen with knowledge of side information at the receiver agent.
\end{IEEEproof}
\begin{pa}
Note that the above composition rule also holds for a one-sided multi-round model in which only one agent shares data for a fixed number of rounds and a net distortion constraint over all rounds. 
\end{pa}
\begin{pa}
We note that composition here focuses on taking a total leakage budget and assigning to optimally to each round; in contrast, composition in differential privacy shows that privacy risk is additive when two different mechanisms are used sequentially on the data. Such a composition rule is generally not straightforward to show for mutual information based metrics.
\end{pa}

{\subsection{Interaction Reduces Leakage: Illustration}}
A natural question in the interactive setting is to understand whether multiple rounds can reduce leakage of the private variables while achieving the desired distortion. In general, it is unclear whether interaction would reduce leakage relative to a one-shot setting. We now present an example for which interaction helps. {To make such a comparison, one could compare the leakage of a specific transmitter agent at the other receiver agent over one round with that over multiple rounds such that in both cases the total number of rounds  culminate at the same receiver agent, i.e., the agent at which a certain level of leakage and distortion is desired. However, depending on whether one chooses odd or even number of rounds, the transmitter agent \textit{need not} be the same for both cases if one were to ensure that the receiver agent is the same. Specifically, if we compare the leakage of a single-round from agent A to agent B against the leakage over two rounds, for the one-shot communications agent A initiates the data sharing. On the other hand, for the two round case the interaction is initiated at agent B such that the second round terminates at agent B, thereby allowing us to compare the one-round leakage of agent A's private data at agent B with that for the two-round interaction setup. We remark that a similar comparison of rate reduction for interactive function computation is developed by Ma \textit{et al.} in  \cite{MA2012}.}

We note that our example is similar to the one in \cite{MA2012} wherein Ma \textit{et. al.} consider an interactive source coding problem for sources $(X_1,X_2)$ at the two agents, i.e., without private data $(Y_1,Y_2)$ and with constraints on coding rate in place of leakage. However, it is not clear the optimal mechanisms for the rate-distortion problem hold when minimizing leakage of $(Y_1,Y_2)$. In fact, one needs to evaluate the optimal mechanism for the problem at hand in each round due to the presence of private side information at each agent and the leakage function being minimized; we detail these computations below.

We consider binary random variables $X_1$, $X_2$, $Y_1$, $Y_2$ such that $(X_1,X_2)$ is modeled as doubly symmetric binary source with parameter $p$, i.e., $(X_1,X_2)\sim \text{DSBS}(p)$,  with $P_{X_1,X_2}(0,0) = P_{X_1,X_2}(1,1) = \frac{1-p}{2}$ and $P_{X_1,X_2}(1,0) = P_{X_1,X_2}(0,1) = \frac{p}{2}$. Furthermore, $(X_1,Y_1)$ and $(X_2,Y_2)$ are correlated as follows: $Y_1=X_1\oplus Z_1$ and $Y_2=X_2\oplus Z_2$ where $Z_i\sim \text{Ber}(p)$ for $i=1,2$, and $Z_1$ and $Z_2$ are independent of $X_1$ and $X_2$, respectively. We let {$d_2=0$} and consider an erasure distortion measure {$d_1(\cdot,\cdot)$} as:
\begin{equation} \label{eq:erasuredist}
d_1(x_1,\hat{x}_1) = \begin{cases} 
      0, & \textrm{ if $\hat{x}_1=x_1$} \\
      1,& \textrm{ if $\hat{x}_1=e$} \\
      \infty, & \textrm{  if $\hat{x}_1=1-x_1$}.\\
   \end{cases} 
   \end{equation}
\textit{One-round sum leakage $L^A_{sum,1}$}: We first compute the sum leakage $L^A_{sum,1}$ for a one round interaction starting from agent A. Note that in this case even though B does not share data, by definition, the sum leakage $L^A_{sum,1}$ includes the leakage of $Y_2$ at A. {In Appendix B}, we show that 
{
   \begin{align}
 L^A_{\text{sum},1}(D_1,0)=&2- [(1-D_1)H(p)+(1+D_1)H(2p(1-p))].  \label{eq:ErasureLeak}
   \end{align}}
\vspace{-0.25in}

For the classical source coding problem with the same distribution defined above for $(X_1, X_2)$ and  functional {$d_1(\cdot,\cdot)$} in (\ref{eq:erasuredist}), the optimal $P_{U_1|X_1}$ minimizing the Wyner-Ziv rate-distortion function $I(X_1;U_1|X_2)$ is well known\cite{ElGamal2011}. However, it is not clear \textit{a priori} that the same transition probability distribution will also minimize the leakage $I(Y_1;U_1|X_2)$ in the presence of private features at both agents. {In Appendix B}, we prove that $I(Y_1;U_1,X_2)$ is indeed minimized by the same distribution that minimizes $I(X_1;U_1|X_2)$. This is also a result of independent interest.

\textit{Two-round sum leakage $L^B_{sum,2}$}: We now compute the sum leakage $L^B_{\text{sum},2}$ for a two-round interaction starting from agent B in round 1 and returning from A to B in round 2. 
Let $U^n_1$ denote the output of the mapping in round 1 from B to A and $U^n_2$ denotes the output of mapping in round 2 from A to B. We will explicitly construct a mechanism pair $(P_{U_1|X_2}, P_{U_2|X_1,U_1})$ and $\hat{X}_1$ which leads to an admissible tuple $(L_1,L_2,D)$. Let $P_{U_1|X_2}$ be binary symmetric channel with crossover probability $\alpha$, i.e., {$P_{U_1|X_2} = \text{BSC}(\alpha)$}. We choose the conditional pmf $P_{U_2|X_1,U_1}$ as given in Table \ref{table:questions} and let $\hat{X}_1=U_2$.
\begin{table}[h]
\caption{Conditional Distribution $P_{U_2|X_1,U_1}$}
\label{table:questions}
\centering
\begin{tabular}{|c|c|c|c|}
\hline
$P_{U_2|X_1,U_1}$ & $u_2=0$ & $u_2=e$ & $u_2=1$ \\ \hline
$x_1=0,u_1=0$     &  $1-\beta$       &  $\beta$        &  0        \\ \hline
               $x_1=1,u_1=0$   &     0    &   1      &  0       \\ \hline
              $x_1=0,u_1=1$      &    0     &  1       & 0        \\ \hline
                  $x_1=1,u_1=1$  &      0   & $\beta$        &   $1-\beta$         \\ \hline
\end{tabular}\vspace{-0.3cm}
\end{table}

For a given value for the DSBS parameter, $p$, there are several values of $(\alpha, \beta)$ pair such that $L^B_{\text{sum},2}\leq L^A_{\text{sum},1} $. For example, for  $p=0.03$, $\alpha =0.35 $, and $\beta =0.55 $, {$L^B_{\text{sum},2}(D_1,0)$} is \vspace{-0.2cm}
\begin{flalign}
\label{2round}
I(Y_2;U_1,X_1)+I(Y_1; U_2|U_1, X_2)=1.1876
\end{flalign}
and the corresponding distortion is {$D_1 = \mathbb{E}(d_1(X_1,\hat{X_1})) = 0.8116 $}. 
By computing $L^A_{\text{sum},1}$ and comparing it with (\ref{2round}) for the same distortion, we have {$L^A_{\text{sum},1}(0.8116,0)=1.3832$}. Thus, interaction reduces leakage. 

In \cite{MA2012}, using the same $P_{U_1|X_2}$, $P_{U_2|X_1,U_1}$ and $\hat{X}_1$ as described above, Ma \textit{et. al.} show that interaction reduces the sum-rate over two rounds relative to one round for  specific values of $p$, $\alpha$, and $\beta$. However, as discussed earlier, it wasn't clear whether the same parameters in \cite{MA2012} also reduce leakage of correlated hidden variables $(Y_1, Y_2)$ in our problem. We have verified that for different value of $\alpha$ and $\beta$ including those in \cite{MA2012}, the two-round sum leakage is smaller than the one-round leakage. 

 \section{When Does Interaction help?}
 \label{sec:sec3}
An important question to address in the interactive setting is whether interaction actually reduces leakage relative to a one-round mechanism. 
In this section, we introduce a test for checking when multiple rounds of interaction help. 
Our approach is modeled along the lines of the method in \cite{MA2013} by Ma \textit{et al.} in which an interactive source coding problem is considered. However, since our source model includes a pair of public and private variables at each agent, we extend the methods in \cite{MA2013} to the problem setting at hand. 
The characterization of $L^A_{\text{sum},K}$ in (\ref{eq:ksum}) does not give us any bounds on the rate of convergence to $L_{\text{sum},\infty}$ for a given distribution $P_{X_1,Y_1,X_2,Y_2}$. { Thus, as in \cite{MA2013}, we use the fact that the sum-leakage function depends on the source distribution only via marginal distributions $P_{X_1,Y_1|X_2}$ and $P_{X_2,Y_2|X_1}$ and characterize the convergence of $L_{\text{sum},K}$ to $L_{\text{sum},\infty}$ for a set of source distributions with the same marginals; this in turn allows us to identify three conditions on the sum-leakage function required for interaction to reduce leakage. }

\sloppy Without loss of generality, let agent $A$ initiate a $K$-round interaction. The goal is to characterize the family of source distributions for which interaction helps. To this end, we define "leakage reduction" functions $\eta^A_K(P_{X_1,Y_1,X_2,Y_2}, D_1,D_2)$ and $\eta^B_K(P_{X_1,Y_1,X_2,Y_2}, D_1,D_2)$ as follows.
\begin{defn}
\label{defn:defn1}
The leakage reduction over $K$ rounds initiated at agent A is defined as
\begin{flalign}
\label{eq:reduction}
&\eta^A_K(P_{X_1,Y_1,X_2,Y_2}, D_1,D_2):= H(Y_1)+H(Y_2) - L^A_{sum, K}(D_1,D_2)\notag\\
&=\max_{P_{U^K|X_1,Y_1,X_2,Y_2}\in \mathcal{P}^A_K}[H(Y_1|U^K,X_2)+H(Y_2|U^K,X_1)].
\end{flalign}
For a $K$-round interaction initiated at agent B, the corresponding leakage reduction function is 
\begin{flalign*}
\label{eq:reductionB}
\eta^B_K(P_{X_1,Y_1,X_2,Y_2}, D_1,D_2):= H(Y_1)+H(Y_2) - L^B_{sum, K}(D_1,D_2)\notag
\end{flalign*}
\begin{flalign}
=\max_{P_{U^K|X_1,Y_1,X_2,Y_2}\in \mathcal{P}^B_K}[H(Y_1|U^K,X_2)+H(Y_2|U^K,X_1)].
\end{flalign}
\end{defn}

Note that $\eta^A_K(P_{X_1,Y_1,X_2,Y_2}, D_1,D_2)$ depends on the distributions $P_{X_1,Y_1|X_2}$ and {$P_{X_2,Y_2|X_1}$}.  Evaluating $\eta^A_K$ is equivalent to evaluating  $L^A_{\text{sum}, K}$.  Definition \ref{defn:defn1} enables us to characterize the properties of  $\eta_{\infty} = \lim_{K\to\infty}\eta^A_K$ which then gives us $L^A_{\text{sum}, \infty} = H(Y_1)+H(Y_2) - \eta_{\infty}$.  The goal is to determine source distributions for which $\eta_{\infty}\geq \eta_{0}$  where $\eta_0$ is the leakage reduction the absence of interaction. When $K=0$, we have $L^A_{\text{sum},0}=L^B_{\text{sum},0}=L_{\text{sum},0} = I(Y_1;X_2)+I(Y_2;X_1)$ and $\eta_0 = H(Y_1|X_2)+ H(Y_2|X_1)$.

For a given source, since it is generally not possible to precisely determine the rate of convergence of $L^A_{\text{sum},K}$ to $L_{\text{sum},\infty}$, we focus, as in \cite{MA2013}, on determining the set of source distributions for which $L_{\text{sum},K}$ is strictly decreasing. This leads us to define the set of structured neighborhoods of $P_{X_1,Y_1,X_2,Y_2}$, i.e.,  a collection of all joint distribution $P'_{X_1,Y_1,X_2,Y_2}$ that have the same marginal $P_{X_2,Y_2|X_1}$ as follows. 
\begin{defn}
The marginal perturbation set $\mathcal{P}_{X_2,Y_2|X_1}$  for a given joint distribution $P_{X_1,Y_1,X_2,Y_2}$ is defined as 
\begin{flalign}
\mathcal{P}_{X_2,Y_2|X_1}(P_{X_1,Y_1,X_2,Y_2})=&\{P'_{X_1,Y_1,X_2,Y_2}: P'_{X_1,Y_1,X_2,Y_2}<< P_{X_1,Y_1,X_2,Y_2}, P'_{X_2,Y_2|X_1}=P_{X_2,Y_2|X_1}\}
\label{eq:MargPertSet}
\end{flalign}
\end{defn}
where $``<<"$ is majorizing operator. One can similarly define $\mathcal{P}_{X_1,Y_1|X_2}(P_{X_1,Y_1,X_2,Y_2})$.

\begin{pa}
 Note that $\mathcal{P}_{X_2, Y_2|X_1}(P_{X_1,Y_1,X_2,Y_2})$ and  $\mathcal{P}_{X_1, Y_1|X_2}(P_{X_1,Y_1,X_2,Y_2})$ are nonempty sets as they contain $P_{X_1,Y_1,X_2,Y_2}$. Furthermore,  for all $P_{X_1,Y_1,X_2,Y_2}$, $\mathcal{P}_{X_2, Y_2|X_1}(P_{X_1,Y_1,X_2,Y_2})$ and $\mathcal{P}_{X_1, Y_1|X_2}(P_{X_1,Y_1,X_2,Y_2})$ are convex sets of $P_{X_1,Y_1,X_2,Y_2}$.
\end{pa}

Recalling that $\eta^A_{K}$ and $\eta^B_{K}$ only depend on $\mathcal{P}_{X_2, Y_2|X_1}$ and $\mathcal{P}_{X_1, Y_1|X_2}$, as a first step towards characterizing $\eta_{\infty}$, we focus only on the set $\mathcal{P}_{X_1,Y_1,X_2,Y_2}$ of source distributions that is closed with respect to marginal perturbations and define it as follows. 
\begin{defn}
A family of joint distributions $\mathcal{P}_{X_1,Y_1,X_2,Y_2}$ is marginal-perturbation-closed if for all  $P_{X_1,Y_1,X_2,Y_2}\in\mathcal{P}_{X_1,Y_1,X_2,Y_2}$, $\mathcal{P}_{X_2, Y_2|X_1}\cup\mathcal{P}_{X_1, Y_1|X_2}\subseteq \mathcal{P}_{X_1,Y_1,X_2,Y_2}$.
\end{defn}

 To characterize $\eta_{\infty}$, we define the following family of functionals.
\begin{defn}
\label{defn:major}
$\eta_0$-majorizing family of functionals $\mathcal{F}_{D}(\mathcal{P}_{X_1,Y_1,X_2,Y_2})$ is the set of all functionals $\eta: \mathcal{P}_{X_1,Y_1,X_2,Y_2}\times\mathcal{D}^2\rightarrow\mathbb{R}$ satisfying
\begin{enumerate}[{(1)}]
\item For all ${P}_{X_1,Y_1,X_2,Y_2}\in\mathcal{P}_{X_1,Y_1,X_2,Y_2}$ and $(D_1,D_2)\in\mathcal{D}^2$, $\eta({P}_{X_1,Y_1,X_2,Y_2},D_1,D_2)\geq \eta_0({P}_{X_1,Y_1,X_2,Y_2},D_1,D_2)$.
\item For all ${P}_{X_1,Y_1,X_2,Y_2}\in\mathcal{P}_{X_1,Y_1,X_2,Y_2}$, $\eta$ is concave on $\mathcal{P}_{X_2,Y_2|X_1}$.
\item For all ${P}_{X_1,Y_1,X_2,Y_2}\in\mathcal{P}_{X_1,Y_1,X_2,Y_2}$, $\eta$ is concave on $\mathcal{P}_{X_1,Y_1|X_2}$.
\end{enumerate}
\end{defn}

To characterize the properties of $\eta_{\infty}$ we need to establish the relationship between  $(k-1)$-round interactive mechanism and $k$-round interactive mechanism. Intuitively speaking, to construct a  $k$-round interactive mechanism, we first pick $U_1$, and then for each realization of $U_1= u_1$, we construct the remaining by considering it as a $(k-1)$-round initiated at agent B but with the distribution $P_{X_1,Y_1,Y_1,Y_2|U_1 = u_1}\in \mathcal{P}_{X_2,Y_2|X_1}(P_{X_1,Y_1,X_2,Y_2})$. As a result, for the remaining $(k-1)$ rounds, the distortion vector will be $(D'_1,D'_2)_{u_1}$ such that $\sum_{u_1}(D'_1,D'_2)_{u_1}P_{U_1}(u_1) = (D_1,D_2)$.  The distortion vector $(D'_1,D'_2)_{u_1}$  for each realization $U_1 = u_1$ in $(k-1)$-round interactive subproblem will, in general,  be different form ($D_1,D_2)$. The following lemma will be used in determining the $\eta_{\infty}$.
\begin{lm}
\label{lm:lm1}
\begin{enumerate}[{(1)}]
\item For all $k\in \mathbb{Z}^+$ and $P_{X_1,Y_1,X_2,Y_2}\in \mathcal{P}_{X_1,Y_1,X_2,Y_2}$, we have
\begin{flalign}
&\eta^A_k(P_{X_1,Y_1,X_2,Y_2},D_1,D_2) =\notag\\
 &\max_{P(U_1|X_1)}\left\{\max_{ \substack {\forall u_1\in \mathcal{U}_1, (D'_1,D'_2)_{u_1}\in\mathcal{D}^2:\\ E((D'_1,D'_2)_{u_1})\leq (D_1,D_2)}}\Big\{\sum_{u_1\in \mathcal{U}_1}P(u_1)\eta^B_{k-1}(P_{X_1,Y_1,X_2,Y_2|u_1}, (D'_1,D'_2)_{u_1})\Big\}\right\}.
\end{flalign}
\item For all $k\in \mathbb{Z}^+$ and all $(q_{X_1,Y_1,X_2,Y_2},D_1, D_2)\in  \mathcal{P}_{X_1,Y_1,X_2,Y_2}\times \mathcal{D}^2$, $\eta^A_k$ is concave on $\mathcal{P}_{X_2,Y_2|X_1}\times\mathcal{D}^2$.
\item For all $k\in \mathbb{Z}^+$ and all $(q_{X_1,Y_1,X_2,Y_2},D_1, D_2)\in  \mathcal{P}_{X_1,Y_1,X_2,Y_2}\times \mathcal{D}^2$, if $\eta:\mathcal{P}_{X_1,Y_1,X_2,Y_2}\times \mathcal{D}^2\rightarrow\mathbb{R}$ is concave on $\mathcal{P}_{X_2,Y_2|X_1}\times\mathcal{D}^2$ and if for all  $(P_{X_1,Y_1,X_2,Y_2},D_1,D_2)\in \mathcal{P}_{X_2,Y_2|X_1}(q_{X_1,Y_1,X_2,Y_2})\times\mathcal{D}^2$, $\eta^B_{k-1}(P_{X_1,Y_1,X_2,Y_2},D_1,D_2)\leq \eta(P_{X_1,Y_1,X_2,Y_2},D_1,D_2)$, then for all $(P_{X_1,Y_1,X_2,Y_2},D_1,D_2)\in \mathcal{P}_{X_2,Y_2|X_1}(q_{X_1,Y_1,X_2,Y_2})\times\mathcal{D}^2$, $\eta^A_{k}(P_{X_1,Y_1,X_2,Y_2},D_1,D_2)\leq \eta(P_{X_1,Y_1,X_2,Y_2},D_1,D_2)$.
 \end{enumerate}
\end{lm}
\begin{IEEEproof}
The proof is very similar to Lemma 1 in \cite{MA2013} and we provide a sketch below. 
\begin{enumerate}[{(1)}]
\item  For all $k\in \mathbb{Z}^+$ and $P_{X_1,Y_1,X_2,Y_2}\in \mathcal{P}_{X_1,Y_1,X_2,Y_2}$
\begin{flalign}
&\eta^A_K(P_{X_1,Y_1,X_2,Y_2}, D_1,D_2)\notag\\
&=\max_{P_{U^K|X_1,Y_1,X_2,Y_2}\in \mathcal{P}^A_K}\Big[H(Y_1|U^K,X_2)+H(Y_2|U^K,X_1)\Big]\label{eq:s0}\\
& =\max_{P_{U_1|X_1}}\left\{ \max_{\substack {P_{U^{K}_2|X_1,Y_1,X_2,Y_2,U_1}: \\ P_{U_1|X_1} P_{U^{K}_2|X_1,Y_1,X_2,Y_2,U_1}\in \mathcal{P}^A_K}}\Big[H(Y_1|U^K,X_2)+H(Y_2|U^K,X_1)\Big]\right\}\label{eq:s1} \\       
& =\max_{P_{U_1|X_1}}\Bigg\{\max_{ \substack {\forall u_1\in \mathcal{U}_1, (D'_1,D'_2)_{u_1}\in\mathcal{D}^2: \\E((D'_1,D'_2)_{u_1})\leq (D_1,D_2)}}\Bigg\{\sum_{u_1}P_{U_1}(u_1)\Bigg\{ \max_{\substack {P_{U^{K}_2|X_1,Y_1,X_2,Y_2,U_1}: \\ P_{U_1|X_1} P_{U^{K}_2|X_1,Y_1,X_2,Y_2,U_1}\in \mathcal{P}^A_K}}\notag\\
&\Big[H(Y_1|U^K_2,X_2,U_1=u_1)+H(Y_2|U^K_2,X_1,U_1 = u_1)\Big]\Bigg\}\Bigg\}\Bigg\}\label{eq:s2}
\end{flalign}
\begin{flalign}
 &=\max_{P(U_1|X_1)}\left\{\max_{ \substack {\forall u_1\in \mathcal{U}_1, (D'_1,D'_2)_{u_1}\in\mathcal{D}^2: \\ E((D'_1,D'_2)_{u_1})\leq (D_1,D_2)}}\left\{\sum_{u_1\in \mathcal{U}_1}P(u_1)\eta^B_{k-1}(P_{X_1,Y_1,X_2,Y_2|u_1}, (D'_1,D'_2)_{u_1})\right\}\right\}\label{eq:s3}
\end{flalign}
where $\mathcal{P}^A_K$ in (\ref{eq:s0}) is as defined in (\ref{prob:probspace}), (\ref{eq:s1}) results from expanding the joint distribution, (\ref{eq:s2}) results from:  (a) replacing the overall distortion constraints by conditional distortion constraints for all $u_1\in\mathcal{U}_1$; (b) using law of total conditional entropy in addition to the fact that conditioned on $U_1=u_1$, $H(Y_1|U^K_2,X_2,U_1=u_1)+H(Y_2|U^K_2,X_1,U_1 = u_1)$ only depends on $P_{U^{K}_2|X_1,Y_1,X_2,Y_2,U_1}$, and (c) for a fixed $P_{U_1|X_1}$, conditioned on $U_1=u_1$, $P_{U_1|X_1}P_{U^{K}_2|X_1,Y_1,X_2,Y_2,U_1}\in\mathcal{P}^A_K(P_{X_1,Y_1,X_2,Y_2|U_1}) $ if and only if  $P_{U^{K}_2|X_1,Y_1,X_2,Y_2,U_1}\in\mathcal{P}^B_{K-1}(P_{X_1,Y_1,X_2,Y_2|U_1}) $.
\item  For all $k\in \mathbb{Z}^+$ and all $(q_{X_1,Y_1,X_2,Y_2},D_1, D_2)\in  \mathcal{P}_{X_1,Y_1,X_2,Y_2}\times \mathcal{D}^2$, consider two arbitrary distributions $P^1_{X_1,Y_1,X_2,Y_2}, P^2_{X_1,Y_1,X_2,Y_2}\in \mathcal{P}_{X_2,Y_2|X_1} $ and distortion vectors $D^1 = (D^1_1,D^1_2), D^2= (D^2_1,D^2_2) \in\mathcal{D}^2$. For every $\lambda\in(0,1)$, define $P^3_{X_1,Y_1,X_2,Y_2} = \lambda P^1_{X_1,Y_1,X_2,Y_2}+\bar\lambda P^2_{X_1,Y_1,X_2,Y_2}$ and $D^3 = \lambda D^1  +\bar\lambda D^2$. We show that $\eta^A_k(P^3_{X_1,Y_1,X_2,Y_2},D^3)\geq \lambda \eta^A_k(P^1_{X_1,Y_1,X_2,Y_2},D^1)+\bar\lambda\eta^A_k(P^2_{X_1,Y_1,X_2,Y_2},D^2)$. Define an auxiliary random variable $V \in \mathcal{U}_1\times\{1,2\}$ such that $P_V(u_1,2)=\bar\lambda P_{U^2_1}(u_1)$ and $P_V(u_1,1)=\lambda P_{U^1_1}(u_1)$ where $ P_{U^1_1},  P_{U^2_1}$ are distributions that maximize (\ref{eq:reduction}) for distributions $P^1_{X_1,Y_1,X_2,Y_2}, P^2_{X_1,Y_1,X_2,Y_2}$, respectively. {According to part (1)} of Lemma \ref{lm:lm1}, we have 
\begin{flalign}
&\lambda \eta^A_k(P^1_{X_1,Y_1,X_2,Y_2},D^1)+\bar\lambda\eta^A_k(P^2_{X_1,Y_1,X_2,Y_2},D^2)\notag\\
&= \lambda\sum_{u_1}{P_{U^1_1}(u_1)\eta^B_{k-1}(P^1_{X_1,Y_1,X_2,Y_2|u_1}, (D^1_1,D^1_2)_{u_1}))}\notag\\
&+\bar\lambda\sum_{u_1}{P_{U^2_1}(u_1)\eta^B_{k-1}(P^2_{X_1,Y_1,X_2,Y_2|u_1}, (D^2_1,D^2_2)_{u_1}))}\notag\\
&=\sum_{\substack{V,\\i=1,2}}{P_V({u_1,i})\eta^B_{k-1}(P^i_{X_1,Y_1,X_2,Y_2|u_1}, (D^i_1,D^i_2)_{u_1}))}\leq \eta^A_k(P^3_{X_1,Y_1,X_2,Y_2},D^3).
\end{flalign}
\item By definition, we can write $\eta^A_k(P_{X_1,Y_1,X_2,Y_2},D_1,D_2)$ as
\begin{flalign}
&\max_{P_{U_1|X_1}}\left\{\max_{ \substack {\forall u_1\in \mathcal{U}_1, (D'_1,D'_2)_{u_1}\in\mathcal{D}^2:\\ E((D'_1,D'_2)_{u_1})\leq (D_1,D_2)}}\Big\{\sum_{u_1\in \mathcal{U}_1}P(u_1)\eta^B_{k-1}(P_{X_1,Y_1,X_2,Y_2|u_1}, (D'_1,D'_2)_{u_1})\Big\}\right\}\\
 \leq&\max_{P_{U_1|X_1}}\left\{\max_{ \substack {\forall u_1\in \mathcal{U}_1, (D'_1,D'_2)_{u_1}\in\mathcal{D}^2:\\ E((D'_1,D'_2)_{u_1})\leq (D_1,D_2)}}\Big\{\sum_{u_1\in \mathcal{U}_1}P(u_1)\eta(P_{X_1,Y_1,X_2,Y_2|u_1}, (D'_1,D'_2)_{u_1})\Big\}\right\}\\
 \leq&\eta(P_{X_1,Y_1,X_2,Y_2})\label{eq:s4}
\end{flalign}
where (\ref{eq:s4}) is due the fact that $\eta$ is concave on $\mathcal{P}_{X_2,Y_2|X_1}\times\mathcal{D}^2$ and Jensen's inequality.
 \end{enumerate}
\end{IEEEproof}
\begin{pa}
\label{remark:3}
By reversing the roles of agents A and B in Lemma \ref{lm:lm1}, one can prove the same lemma for agent B.
\end{pa}
Using Lemma \ref{lm:lm1}, we now have the following theorem relating $\eta_{\infty}$ to the set $\mathcal{F}_{\mathbf{\mathcal{D}}}({\mathcal{P}_{X_1,Y_1,X_2,Y_2}})$.
\begin{thm}
\label{thm:thmleast}
$\eta_{\infty}({P}_{X_1,Y_1,X_2,Y_2}, D_1,D_2)\in \mathcal{F}_{\mathbf{\mathcal{D}}}({\mathcal{P}_{X_1,Y_1,X_2,Y_2}})$ and is its least element. \end{thm}
\begin{IEEEproof}
 We show that $\eta_{\infty}$ satisfies all three conditions in Definition \ref{defn:major} as follows:
\begin{enumerate}[{(1)}]
\item {Condition (1)} in Definition  \ref{defn:major} is satisfied since $L_{\text{sum},\infty}\leq L_{\text{sum},0}$, due to (\ref{eq:AA}) in Lemma \ref{lm:lm0}. 
\item {Condition (2)} in Definition  \ref{defn:major} is satisfied from step 2 of Lemma \ref {lm:lm1}.
\item {Condition (3)} in Definition \ref{defn:major} is satisfied due to Remark \ref{remark:3}.
\end{enumerate}
To show that $\eta_{\infty}$ is the smallest element of $\mathcal{F}_{\mathbf{\mathcal{D}}}({\mathcal{P}_{X_1,Y_1,X_2,Y_2})}$: we need to show that for all $\eta\in \mathcal{F}_{\mathbf{\mathcal{D}}}({\mathcal{P}_{X_1,Y_1,X_2,Y_2}})\times\mathcal{D}^2$,  for all ${P}_{X_1,Y_1,X_2,Y_2}\in\mathcal{P}_{X_1,Y_1,X_2,Y_2}$, and, for all $k$, $\eta^A_k({P}_{X_1,Y_1,X_2,Y_2})\leq\eta({P}_{X_1,Y_1,X_2,Y_2})$ and $\eta^B_k({P}_{X_1,Y_1,X_2,Y_2})\leq\eta({P}_{X_1,Y_1,X_2,Y_2})$. Using induction on $k$ and conditions (2) and (3) of Lemma \ref{lm:lm1}, we have that for all $k$, $\eta^A_k\leq\eta$, and thus, $\eta_{\infty}$ is the least element of $\mathcal{F}_{\mathbf{\mathcal{D}}}({\mathcal{P}_{X_1,Y_1,X_2,Y_2}})$.
\end{IEEEproof}
Having determined the conditions under which $\eta_{\infty}$ is the smallest value of $\mathcal{F}_{\mathbf{\mathcal{D}}}$, i.e., the conditions under which interaction helps, we can now identify the conditions under which interaction does not help.
\begin{thm}
\label{thm:cond}
The following equivalent conditions establish when interaction does not help.
\begin{enumerate}[{(1)}]
\item For all ${P}_{X_1,Y_1,X_2,Y_2}\in\mathcal{P}_{X_1,Y_1,X_2,Y_2}$ and $D=(D_1,D_2)\in\mathcal{D}^2$, $\eta^A_k({P}_{X_1,Y_1,X_2,Y_2},D) = \eta_{\infty}({P}_{X_1,Y_1,X_2,Y_2},D)$.
\item For all ${P}_{X_1,Y_1,X_2,Y_2}\in\mathcal{P}_{X_1,Y_1,X_2,Y_2}$ and $D=(D_1,D_2)\in\mathcal{D}^2$, $\eta^A_k({P}_{X_1,Y_1,X_2,Y_2},D) = \eta^B_{k+1}({P}_{X_1,Y_1,X_2,Y_2},D)$.
\item For all ${P}_{X_1,Y_1,X_2,Y_2}\in\mathcal{P}_{X_1,Y_1,X_2,Y_2}$ and $D=(D_1,D_2)\in\mathcal{D}^2$,  $\eta^A_k$ is concave on $\mathcal{P}_{X_1,Y_1|X_2}(P_{X_1,Y_1,X_2,Y_2})\times \mathcal{D}^2$.
\end{enumerate}
\end{thm}
\begin{IEEEproof}
{Condition (1) implies condition (2)} since from Lemma \ref{lm:lm0}, $\eta^A_k\leq\eta^B_{k+1}\leq\eta_{\infty}$. {Condition (2) implies condition (3)} due to Remark \ref{remark:3}. {Condition (3) implies condition (1)} can be shown by using {part (2)} in Lemma \ref{lm:lm1} along with the fact that $\eta^A_{k}\geq \eta_0$, which leads to $\eta^A_{k}\in \mathcal{F}_{\mathbf{\mathcal{D}}}({\mathcal{P}_{X_1,Y_1,X_2,Y_2})}$.  From Theorem \ref{thm:thmleast}, since $\eta_{\infty}$ is the least element of $ \mathcal{F}_{\mathbf{\mathcal{D}}}({\mathcal{P}_{X_1,Y_1,X_2,Y_2})}$ we have $\eta^A_k\geq\eta_{\infty}$. Therefore, $\eta^A_k = \eta_{\infty}$.
\end{IEEEproof}

\section{Interactive Privacy Funnel: Private Interactive Mechanisms Under Log-Loss Distortion}
\label{sec:sec6}
 Logarithmic loss is a widely used penalty function in machine learning theory and prediction and it is a natural loss criterion in scenarios where reconstructions are allowed to be soft, i.e., they can be probability measures instead of deterministic decision values. We now derive the leakage-distortion region under log-loss distortion. 

Formally, for a random variable $X\in\mathcal{X}$ and its reproduction alphabet $\hat{\mathcal{X}}$ as the set of probability measures  on $\mathcal{X}$, the log-loss distortion is defined as 
\begin{equation}
d(x,\hat{x}) = \log(\frac{1}{\hat{x}(x)}).
\end{equation}
\subsection{Leakage-distortion region for log-loss distortion}
\begin{thm}
For the $K$-round interaction mechanism the leakage-distortion region under log-loss distortion, set of all tuples $(L_1, D_1, L_2,D_2)$ is given by:
 \begin{flalign}
(L_1,L_2,D_1,D_2) : &\ L_1\geq I(Y_1; U_1,\dots, U_K, X_2),\notag \\
&\ L_2\geq I(Y_2; U_1,\dots, U_K, X_1),\notag \\
&D_1\geq H(X_1|U_1,\dots,U_K,X_2)\notag\\
&D_2\geq H(X_2|U_1,\dots,U_K,X_1).
\label{eq:loglossdist}
\end{flalign}
\label{thm:thm2}
\end{thm}
\vspace{-0.25in}
\begin{IEEEproof}
The distortion bounds in (\ref{eq:loglossdist}) result from applying  $\hat{X}_i = P(X_i=x_i|U_1,\dots,U_K,X_j)$ $i =1,2$, $j \neq i$, to the distortion bounds given by (\ref{eq:LU_def}) in Theorem~\ref{thm: thm1}, to get
\begin{flalign}
&D_i\geq \mathbb{E}(d(X_i,\hat{X}_i)) \notag\\
=& \sum_{x_i,u_1,\dots, u_K}{P(x_i,u_1,\dots, u_K)}\log(\frac{1}{P(x_i|u_1,\dots,u_K,x_j)})= H(X_i|U_1,\dots,U_K,X_j),
 \end{flalign}
 where the summation is over $(x_i, u_1,\dots, u_K)$ since $\hat{X}$ is a function of $(U_1,\dots, U_K)$.
\end{IEEEproof}
\begin{ppp}
For special case, $Y_i = X_i$, $i = 1,2$, we have $L_1(D_1,D_2) = H(Y_1)-D_1$ and $L_2(D_1,D_2) = H(Y_2)-D_2$, i.e., the leakage for each agent is simply the rate-distortion function under log-loss distortion.
\end{ppp}

\begin{figure}[htbp]
	\begin{center}
	\includegraphics[width=8cm,keepaspectratio]{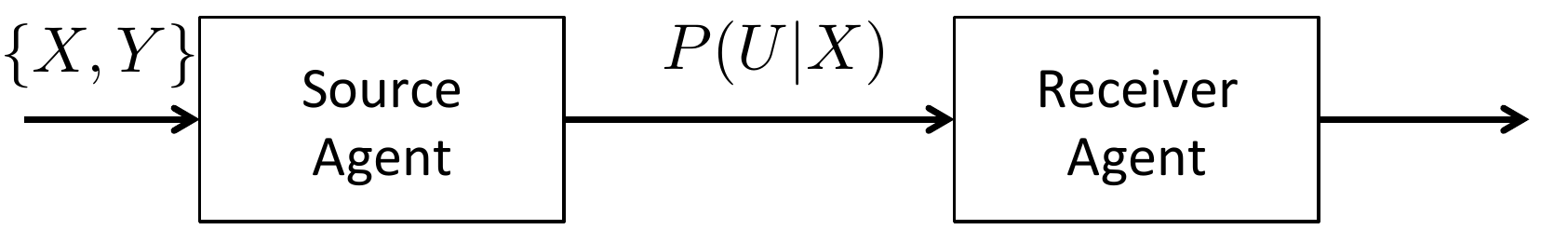}
		\caption{One-way non-interactive mechanism.}
		\label{fig:fig2}
	\end{center}\vspace{0cm}
\end{figure}

For the case $Y_i = X_i$, $i = 1,2$, as explained earlier, the leakage-distortion region is the same as the rate-distortion region. In \cite{Courtade2011}, it is shown that a one-shot scheme achieves the rate-distortion region. In fact, the optimal mapping is a one-shot Wyner-Ziv scheme that each agent uses to share data simultaneously and independently with the other agent. 
\begin{ppp}
For $X_2 = Y_2 = \emptyset$, under log-loss distortion measure and $K=1$, i.e., a non-interactive one-round setting with a single source agent and single receiver agent (see Fig. { \ref{fig:fig2}}), the bounds in Theorem {\ref{thm:thm2}} yields the following optimization problem:
 \begin{align}\label{eq:funnel}
\min_{P_{U|X}: I(X;U)\geq\tau} {I(Y;U)}.
\end{align}
\end{ppp}

In general, when $Y_i \neq X_i$, $i = 1,2$, a one-shot scheme will not achieve the set of all $(L_1, D_1, L_2, D_2)$ tuples in Theorem \ref{thm:thm2}. It is then of interest to understand if interaction reduces leakage, and if so, to determine the optimal set of mechanisms. To this end, we begin by rewriting the distortion bounds in (\ref{eq:loglossdist}) as
 \begin{align}
 \label{eq1:rewrite}
&I(X_1;U_1,\dots,U_K,X_2)\geq\tau_1\\
&I(X_2;U_1,\dots,U_K,X_1)\geq\tau_2.\notag
\end{align}

From (\ref{eq:loglossdist}), computing the $K$-round sum leakage leads to the following optimization problem:
\begin{equation}\label{eq:optimization}
\min_{\left\{  P_{1k},P_{2k}\right\}  _{k=1}^{K/2}}\sum_{i,j=1,i\not =j}%
^{2}I(Y_{i};U_{1,}...,U_{K},X_{j})
\end{equation}
such that for all $i,j=1,2,i\not =j,$%
\begin{equation}
\label{eq:reg}
I(X_{i};U_{1},...U_{K},X_{j})\geq\tau_{i}.
\end{equation}

We refer to the optimization problem in (\ref{eq:optimization}) as an \textit{interactive privacy funnel problem}. The optimization problem (\ref{eq:optimization}) is not convex because of the non-convexity of  the feasible region in (\ref{eq:reg}). One can, however, draw parallels between the above optimization problem and the information bottleneck (IB) problem that Tishby \textit{et al.} introduce in \cite{Tishby2000} in which for a source $(X,Y)$ and an output $U$ such that $Y\leftrightarrow X\leftrightarrow U$ form a Markov source, the goal is to minimize the information shared about $X$ via $U$ while preserving a measure of information about the correlated feature $Y$ via $U$. One can see immediately that the IB problem is a dual of the privacy problem considered here in that the features to be revealed and hidden are swapped. Noting that the IB optimization problem is non-convex, the authors in \cite{Slonim1999} present an agglomerative information bottleneck algorithm that is guaranteed to converge to a local minima. Recently, in \cite{Makhdoumi2014}, Makhdoumi \textit{et al.} also observe parallels between the information bottleneck problem and the single-round version of the problem considered here; i.e., for the case of a one-way non-interactive single source agent and a single receiver agent setup (with no side information at the receiver agent) shown in Fig.~\ref{fig:fig2} and the associated privacy funnel optimization problem in (\ref{eq:funnel}). Furthermore, they apply Slonim's algorithm to their privacy funnel setup to compute a locally optimal mechanism. The optimization we study in (\ref{eq:optimization}) is an interactive version of (\ref{eq:funnel}), and thus, requires generalizing the methods and approaches for the non-interactive case to the interactive setup. In the following subsections we first introduce the information bottleneck problem, the agglomerative information bottleneck algorithm, agglomerative interactive privacy algorithm, using a merge-and-search technique that we introduce to generalize the agglomerative information bottleneck algorithm to the multi-round setting, and show how the presence of side-information in each round is used to generalize the algorithm.

\subsection{Information Bottleneck Problem and Agglomerative Information Bottleneck Algorithm}
Consider the setting in Fig. \ref{fig:fig2} with $X_2 = \emptyset$ and $Y_2 = \emptyset$. The information bottleneck problem seeks to minimize the compression rate between $X$ and $U$, while preserving a measure of the average information between $U$ and some correlated data $Y$ and is given by
\begin{flalign}
\label{eq:IBeq}
&\min_{P_{U|X}: I(Y;U)\geq\tau} {I(X;U)}.
\end{flalign}

In \cite{Tishby2000}, Tishby \textit{et al.} show that it is possible to characterize the general form of a locally optimal solution for the information bottleneck problem in (\ref{eq:IBeq}) and develop an iterative algorithm to do so. Furthermore, for ease of computation, in \cite{Slonim1999}, Slonim \textit{et al.} introduce an agglomerative information bottleneck algorithm which guarantees the identification of at least one locally optimal solution with lower computational complexity than the iterative algorithm \cite{Slonim1999}. The agglomerative algorithm, as the name suggests, involves reducing the cardinality of the auxiliary random variable $U$ iteratively until the constraints on both $X$ and $Y$ are satisfied, in \cite{Slonim1999}, the authors prove that it converges to a local minima of the optimization problem. We adopt this algorithm and generalize it to the interactive setting.

We first briefly outline the agglomerative information bottleneck algorithm which yields a solution to (\ref{eq:IBeq}). The procedure typically starts with the most fine-grained solution where ${\mathcal{U}}=\mathcal{X}$, i.e., each value of $X$ is assigned to a unique singleton cluster in $U$. The idea is to reduce the cardinality of ${\mathcal{U}}$ and consequently reduce $I(X;U)$, by merging two values of $u_i\in {\mathcal{U}}$ and ${u}_j \in{\mathcal{U}}$ such that the new merged random variable $U_{ij}$ is distributed as 
\begin{flalign}
P({u}_{ij}|x) = P(u_i|x) +P(u_j|x)
\end{flalign}

In the $k^{th}$ iteration, the indices $i$ and $j$ are chosen that ${U}^k_{ij}$ satisfies the constraint in (\ref{eq:IBeq}), while $I(X; U^k_{ij})$ is at most as large as $I(X;U^{k-1})$ where $U^{k-1}$ denotes the random variable from the previous iteration. 

In\cite{Makhdoumi2014}, the authors apply the agglomerative information bottleneck algorithm to compute the locally optimal leakage for a desired $\tau$ in (\ref{eq:funnel}). They refer to the optimization problem in (\ref{eq:funnel}) as a privacy funnel problem and the resulting optimization algorithm as greedy algorithm privacy funnel.

As observed in \cite{Makhdoumi2014}, we note that the optimization problem in (\ref{eq:optimization}) as well as (\ref{eq:funnel}) differs from the information bottleneck problem in (\ref{eq:IBeq}) in that the minimization and constraint functions are swapped for the same minimizing argument.
\subsection{Agglomerative Interactive Privacy Algorithm}
\begin{figure}[htbp]
	\begin{center}
	\includegraphics[width=8cm,keepaspectratio]{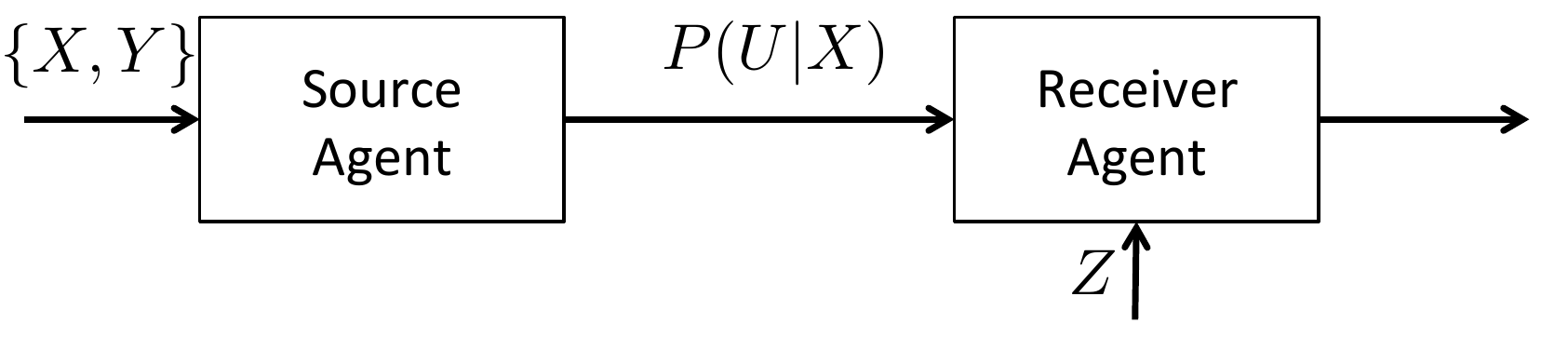}
		\caption{Point to point mechanism with side information}
		\label{fig:fig3}
	\end{center}\vspace{0cm}
\end{figure}

The optimization problem in (\ref{eq:optimization}) is an interactive generalization of the privacy funnel problem in (\ref{eq:funnel}) in which both agents have access to data sources that need to be shared. We now show that the multi-round interaction setup allows a natural generalization of the single round case. To develop such an algorithm, we first consider the single round case with side information at the receiver agent (depicted in Fig. \ref{fig:fig3}). We introduce a \textit{merge-and-search technique} that extends the agglomerative information bottleneck algorithm described earlier to a multivariate setting.

\textit{Merge-and-Search technique}: Consider a one-round setting, i.e., $K=1$ with side information at receiver agent (Fig. \ref{fig:fig3}). Since $I(Y;Z)$ is fixed by joint source distribution, the optimization problem in (\ref{eq:optimization}) can be simplified as 
\begin{equation}
\label{eq:siopt}
\min_{P_{U|X}}I(Y;U, Z)\hspace{0.2cm} \text{s.t.}\hspace{0.2cm}I(X;U, Z)\geq\tau_1\\
\end{equation}
Comparing with (\ref{eq:funnel}), the optimization in (\ref{eq:siopt}) is obtained by replacing $U$ by the tuple $(U, Z)$ and $P_{U|X}$ by $ P_{U, Z|X}= P_{U|X}P_{Z|X}$. Thus, in computing the optimal mechanism, one now needs to consider the pair $(U, Z)$. We iteratively reduce the cardinality of $U$ to reduce $I(Y;U, Z)$ by merging the values of $U$ for each value of $Z$ such that distortion condition in (\ref{eq:siopt}) is satisfied, i.e., in the $k$-th iteration, we choose indices $i$ and $j$ such that $I(X; U^{k}_{ij},Z)\geq \tau_1$ where $U^k_{ij}$ is the resulting from merging $u_i$ and $u_j$ while maximizing $I(Y;U^{k-1}|Z) - I(Y;U^k_{ij}|Z)$ where $U^{k-1}$ is the output of the algorithm in round ${k-1}$. These steps are the basis of our merge-and-search technique that extends ~\cite[Algorithm 1]{Makhdoumi2014} to the more general point-to-point setting with side information at receiving agent. 

Consider the two-round setting in (\ref{eq:optimization}), i.e., $K=2$. We can use the above described merge-and-search technique iteratively to find the mechanism $(P_{11},P_{21})$. In the first round, we have a point-to-point setting with side information $X_2$ for which the distribution $P_{U_1|X_1}$ can be found, as detailed above. In the second round, the cardinality of $U_2$ is reduced to decrease $I(Y_2;U_1,U_2,X_1)$ using $P_{U_1, X_1}$ computed during the first round. This reduction is computed by merging elements of $U_2$ conditioned on $U_1$ and $X_1$.

The steps we outlined above can be extended to find the locally optimal mechanism $\{P_{1i}, P_{2i}\}_{i=1}^{\frac{K}{2}}$ for any $K\geq2$ and is detailed in Algorithm 1. 
\begin{table}
\begin{center}
\label{alg:alg1}
\normalsize
\begin{tabular}{l}
\hline
{\bf Algorithm 1}: Agglomerative Iterative Algorithm\\
\hline
\hspace{0.1cm} For $k=1,\dots, K/2$\\
\hspace{0.1cm} {\bf R(2k-1)}: $\min I(Y_1;X_2, U_1,\dots, U_{2k-2}, U_{2k-1})$ \\
\hspace{1.5cm} over $P_{U_{2k-1}|X_2, U_1,\dots,U_{2k-2}}$\\
\hspace{1.5cm}s.t. $I(X_1;U_{2k_1}|X_2,U_1,\dots, U_{2k-2})\geq\tau_{2k-1}$\\
\hspace{0.3cm} {\bf Input (2k-1):} $P_{X_1,Y_1}$, $P_{U_{2k-2},\dots, U_1, X_1,X_2}$, $\tau_{2k-1}$\\
\hspace{0.7cm} Apply the merge-and-search technique to find local optimum.\\
\hspace{0.3cm} {\bf Output (2k-1):} $P_{U_{2k-1}|X_1, X_2,U_1,\dots, U_{2k-2}}$ \\
\hspace{0.1cm} {\bf R(2k)}: $\min I(Y_2; X_1, U_1,\dots, U_{2k-1}, U_{2k})$ \\
\hspace{1.5cm} over $P_{U_{2k}|X_1, U_1,\dots,U_{2k-1}}$\\
\hspace{1.5cm}s.t. $I(X_2;U_{2k}|X_1,U_1,\dots, U_{2k-1})\geq\tau_{2k}$\\
\hspace{0.3cm} {\bf Input (2k):} $P_{X_2,Y_2}$, $P_{U_{2k-1},\dots, U_1, X_1,X_2}$, $\tau_{2k}$\\
\hspace{0.7cm} Apply the merge-and-search technique to find local optimum.\\
\hspace{0.3cm} {\bf Output (2k):} $P_{U_{2k}|X_1, X_2,U_1,\dots, U_{2k-1}}$\\
\hspace{0.3cm}{\bf Output :} $P_{U_1|X_1},\dots, P_{U_K|U_1,\dots,U_{K-1},X_2}$\\
\hline
\end{tabular}
\end{center}
\end{table}
We note that since at each round the side-information aware locally optimal mechanism is found by the merge-and-search algorithm, the resulting solution over $K$ rounds is also locally optimal. This follows directly from the composition property of the information-theoretic privacy mechanism discussed {{in Section \ref{lb:2A}}}. 
\subsection{Gaussian Sources Under Log-Loss Distortion}
In this section, we prove for Gaussian sources under log-loss distortion one round of interaction suffices. We leverage the results by Tishby \textit{et.al} for the non-interactive case in \cite{Tishby2004} to prove that no interaction is required.
\newtheorem{qw}{Proposition}
\begin{qw}{[ \cite{Tishby2004}, Theorem 1]}
\label{prop:prop1}
Let $(X,Y)$ be jointly Gaussian distributed. Let $U$ be the output of a mapping $P_{U|X}$ such that $Y\leftrightarrow X\leftrightarrow U$ forms a Markov chain. The mapping $P_{U|X}$ that minimizes the following information bottleneck problem for jointly Gaussian sources 
\begin{equation}
\label{eq:IB}
\begin{aligned}
& \underset{P_{U|X}}{\text{min}}
& & I(X;U) \\
& \text{subject to}
& & I(Y;U)\geq\tau.
\end{aligned}
\end{equation}
is Gaussian.
\end{qw}

The above result extends in a straightforward manner to the non-interactive (one-way) privacy funnel setting given by (\ref{eq:funnel}) and we summarize it in the following corollary. The extension is a direct result of the fact that since $X$ and $Y$ are jointly Gaussian, the optimal mapping remains Gaussian even when the objective and constraint functions are swapped in (\ref{eq:funnel}).
\begin{ppp}
\label{Cor:privacy}
For the non-interactive (one-way) single source and single receiver agent setting in Fig. \ref{fig:fig2} with the leakage-distortion tradeoff problem given by (\ref{eq:funnel}), the optimal leakage-minimizing mechanism is Gaussian.
\end{ppp}

As a first step towards establishing optimality of a one-round Gaussian mechanism for the interactive setting, we extend the results in Proposition \ref{prop:prop1} to the case in which the receiver agent, chosen as agent 2 without loss of generality in the non-interactive setting, has side information $Z$ correlated with source date $(X,Y)$.
\begin{lm}
\label{thm:thm5}
Suppose $(X,Y)$ and $(X,Z)$ are jointly Gaussian and let $P_{U|X}$ be a privacy mechanism such that $U\leftrightarrow X\leftrightarrow Z$ forms a Markov chain  (see Fig. \ref{fig:fig3}). The optimal mechanism $P_{U|X}$ minimizing $I(Y;U,Z)$ subject to $I(X;U,Z) \geq \tau$ is Gaussian.
\end{lm}
\begin{IEEEproof}
Define $V = (U,Z)$. Now, consider the following optimization problem 
\begin{equation}
\label{eq:IB1}
\begin{aligned}
& \underset{P_{V|X}}{\text{min}}
& & I(Y;V) \\
& \text{subject to}
& & I(X;V)\geq\tau.
\end{aligned}
\end{equation}. 
From Corollary {\ref{Cor:privacy}}, the optimizing mechanism $P_{V|X}$, and therefore, the output $V$ in (\ref{eq:IB1}) are Gaussian. Thus, since $Z$ is Gaussian, we have that $(U,Z)$ are jointly Gaussian. Note that the mechanisms over which the optimizations are done in Lemma {\ref{thm:thm5}} and (\ref{eq:IB1}) are the same since $P_{V|X} = P_{U,Z|X} = P_{Z|X}P_{U|X}$ for a given source distribution $P_{X,Z}$. 
\end{IEEEproof}

We now use Lemma {\ref{thm:thm5}} to determine the optimal mechanism for the $K$-round interactive mechanism with Gaussian sources and show that one round of interaction suffices. 
\begin{thm}
Consider a two-agent interactive setting with log-loss distortion and jointly Gaussian sources. The optimal leakage-distortion tradeoff region in Theorem \ref{thm:thm2} can be achieved in one round of interaction.
\end{thm}
\begin{IEEEproof}
From Lemma \ref{thm:thm5} we have that {for a Gaussian source transmitting to a receiver agent with jointly Gaussian side information, the optimal mechanism is Gaussian. Since the interactive setting involves a set of $K$ such mechanisms, it is straightforward to see that the tuple $(U_1,\dots,U_K)$ in Theorem \ref{thm:thm2} should also be Gaussian. This in turn implies that one could choose a single Gaussian random variable $U_1$ correlated in such a manner with the public data $X_i$ at the transmit agent $i$ such that the resulting leakage is the same as over $K$ rounds, i.e., one round of interaction suffices.}
\end{IEEEproof}
\subsection{Benefit of Interaction Under Log-Loss Distortion}

In this section, we will show that there exists at least one source for which  multiple rounds of interaction help under log-loss distortion using Theorem \ref{thm:cond}. 

Let $\mathcal{X}_1 =\mathcal{X}_2 = \mathcal{Y}_1=\{0,1\}$ and $\mathcal{Y}_2 = \emptyset$. We choose $(X_1,X_2)\sim \text{DSBS}(p)$, and $Y_1 = X_1+N$, $N\sim \text{Ber}(r)$ such that $X_1$ and $N$ are independent and $r\in(0,1)$. Furthermore, let $P_{X_1,Y_1|X_2}$ be the distribution in Table \ref{table:benefit} and let the distortion pair be $(D_1,D_2) = (\infty,D)$. We focus on a one-round mechanism from agents A to B and show that $\eta^A_1$ is not a concave function of $\mathcal{P}_{X_1,Y_1|X_2}$; since $\mathcal{P}_{X_1,Y_1|X_2}$ is the set of all joint distributions $P_{X_1,Y_1,X_2}$ with the same marginals $P_{X_1,Y_1|X_2}$, to verify concavity, we consider two distributions on $X_2$ such that $P^{(1)}_{X_2}=\text{Ber}(q)$ and $P^{(2)}_{X_2}=\text{Ber}(\bar{q})$ where $\bar{q} = 1-q$ (i.e., we verify condition 3 in  Theorem \ref{thm:cond} which in turn relies on the definition of marginal perturbation sets in (\ref{eq:MargPertSet})). Note that for $P_{X_2} =\frac{P^{(1)}_{X_2}+P^{(1)}_{X_2}}{2} $ we have $X_2 \sim \text{Ber}(\frac{1}{2})$. Since computing the optimal $\eta^A_1$ is not straightforward, we develop upper and lower bounds which allow us to verify the concavity condition. From Theorem \ref{thm:cond}, it is sufficient to show that there exist $P_{X^{(1)}_2}$ and $P_{X^{(2)}_2}$ such that
\begin{flalign}
\label{eq:eta_concavity}
\eta^A_1(P_{X_1,Y_1|X_2}\frac{P_{X^{(1)}_2}+P_{X^{(2)}_2}}{2}, D) < \frac{\eta^A_1(P_{X_1,Y_1|X_2}P_{X^{(1)}_2}, D)+\eta^A_1(P_{X_1,Y_1|X_2}P_{X^{(2)}_2},D)}{2}
\end{flalign}
To this end, we develop upper and lower bounds on the left and right sides of (\ref{eq:eta_concavity}), respectively, and show that there exists jointly distributed sources and a mechanism for which the lower bound is strictly larger than the upper bound. We develop the lower bound in the following lemma and prove the existence of a source and mechanism in the theorem that follows.

\begin{table}[h]
\caption{Conditional Distribution $P_{X_1, Y_1|X_2}$}
\label{table:benefit}
\centering
\begin{tabular}{|c|c|c|c|}
\hline
$P_{X_1,Y_1|X_2}$ & $X_2=0$ & $X_2=1$ \\ \hline
$X_1=0,Y_1=0$     &  $\bar p\bar r $    &  $p\bar r$         \\ \hline
               $X_1 = 0,Y_1=1$   &     $\bar p r $    &   pr      \\ \hline
              $X_1=1,Y_1=0$      &    pr    &  $\bar p r $               \\ \hline
                  $X_1=1,Y_1=1$  &      $ p\bar r $  & $\bar p\bar r $             \\ \hline
\end{tabular}\vspace{0cm}
\end{table}
\begin{lm}
\label{lm:bound}
For $P^{(1)}_{X_2}\sim \text{Ber}(q)$, $P^{(2)}_{X_2}\sim \text{Ber}(\bar{q})$, and $P_{X_1,Y_1|X_2}$ as in Table \ref{table:benefit}, and the distortion function at agent B $H(X_1|X_2,U) = \gamma(p,q,r,\alpha_{2,0}, \alpha_{2,1}) = \bar{q}(\bar{p}\alpha_{2,0}+p\alpha_{2,1}) H(\frac{\bar{p}\alpha_{2,0}}{\bar{p}\alpha_{2,0}+p\alpha_{2,1}})+{q}({p}\alpha_{2,0}+\bar{p}\alpha_{2,1}) H(\frac{{p}\alpha_{2,0}}{{p}\alpha_{2,0}+\bar{p}\alpha_{2,1}})$, we have
\begin{flalign}
\frac{\eta^A_1(P_{X_1,Y_1|X_2}P^{(1)}_{X_2}, D)+\eta^A_1(P_{X_1,Y_1|X_2}P^{(2)}_{X_2},D)}{2} \geq C(p,q,r,\alpha_{2,0}, \alpha_{2,1})
\end{flalign}

where
\begin{flalign}
\label{eq:C}
C(p,q,r,\alpha_{2,0}, \alpha_{2,1}) =&\bar{q}(\bar{p}\bar{\alpha}_{2,0}+p\bar{\alpha}_{2,1})H(r)+q(p\bar{\alpha}_{2,0}+\bar{p}\bar{\alpha}_{2,1})H(r)\notag\\
+&\bar{q}(\bar{p}\alpha_{2,0}+p\alpha_{2,1}) H(\frac{\bar{p}\bar{r}\alpha_{2,0}+pr\alpha_{2,1}}{\bar{p}\alpha_{2,0}+p\alpha_{2,1}})\notag\\
+&q(p\alpha_{2,0}+\bar{p}\alpha_{2,1})H(\frac{pr\alpha_{2,0}+\bar{p}\bar{r}\alpha_{2,1}}{p\alpha_{2,0}+\bar{p}\alpha_{2,1}})
\end{flalign}

for  $0\leq\alpha_{2,0}, \alpha_{2,1}\leq 1$ and 
\begin{equation}
\label{eq:odist}
P_{U|X_1}(u|x_1)=  \begin{bmatrix}
 1-\alpha_{2,0}& 0 \\
0& 1-\alpha_{2,1} \\
\alpha_{2,0} & \alpha_{2,1} 
  \end{bmatrix}.
\end{equation}
\end{lm}
\begin{IEEEproof}
Computing $\eta^A_1(P_{X_1,Y_1,X^{(1)}_2},D)$ simplifies to the following optimization problem:
\begin{equation}
\begin{aligned}
\label{eq:LK1}
& \underset{P_{U|X_1}}{\text{max}}
& & H(Y_1|X^{(1)}_2, U) \\
& \text{subject to}
& & H(X_1|X^{(1)}_2,U)\leq D.
\end{aligned}
\end{equation}

One can obtain a lower bound on $\eta^A_1(P_{X_1,Y_1,X^{(1)}_2},D)$ by computing (\ref{eq:LK1}) for a specific $P_{U|X_1}(u|x_1)$ that we choose as 
\begin{equation}
\label{eq:odist1}
P_{U|X_1}(u|x_1)=  \begin{bmatrix}
 1-\alpha_{2,0}& 0 \\
0& 1-\alpha_{2,1} \\
\alpha_{2,0} & \alpha_{2,1} 
  \end{bmatrix}.
\end{equation}
We observe that for the choice of mechanism in (\ref{eq:odist1}), $H(Y_1|X^{(1)}_2, U)$ is simply $C(p,q,r,\alpha_{2,0}, \alpha_{2,1})$, and thus, we have $\eta^A_1(P_{X_1,Y_1|X_2}P_{X^{(1)}_2}, D)\geq C(p,q,r,\alpha_{2,0}, \alpha_{2,1})$.

Furthermore, from (\ref{eq:C}) observe that $C(p,q,r,\alpha_{2,0}, \alpha_{2,1})=C(p,\bar{q},r, \alpha_{2,1},\alpha_{2,0})$ and $\gamma(p,q,r,\alpha_{2,0}, \alpha_{2,1})=\gamma(p,\bar{q},r, \alpha_{2,1},\alpha_{2,0})$, i.e., $C(\cdot)$ and $\gamma(\cdot)$ are unchanged when the tuple $(q,r,\alpha_{2,0}, \alpha_{2,1})$ is replaced by $(\bar{q},r, \alpha_{2,1},\alpha_{2,0})$. Therefore we have:
\begin{flalign}
\eta^A_1(P_{X_1,Y_1|X_2}P_{X^{(2)}_2}, D) \geq& \max_{\substack{\alpha_{2,0}, \alpha_{2,1}\in[0,1]:\\\gamma(p,\bar{q},r,\alpha_{2,0}, \alpha_{2,1})\leq D}}
C(p,\bar{q},r,\alpha_{2,0}, \alpha_{2,1}) \notag\\
=&\max_{\substack{\alpha_{2,0}, \alpha_{2,1}\in[0,1]:\\\gamma(p,q,r,\alpha_{2,0}, \alpha_{2,1})\leq D}} C(p,q,r, \alpha_{2,1},\alpha_{2,0})\notag\\
\geq& C(p,q,r, \alpha_{2,0},\alpha_{2,1})
\end{flalign}

from which it follows that 
\begin{flalign}
\frac{\eta^A_1(P_{X_1,Y_1|X_2}P_{X^{(1)}_2}, D)+\eta^A_1(P_{X_1,Y_1|X_2}P_{X^{(2)}_2}, D)}{2} \geq & {C(p,q,r,\alpha_{2,0}, \alpha_{2,1})}
\end{flalign}
\end{IEEEproof}
\begin{thm}
\label{thm:benefit}
Under log-loss distortion measure, there exists a joint probability distribution $P_{X_1,Y_1,X_2,Y_2}$, $Y_2 = \emptyset$, and a distortion pair $D = (D_1,D_2) = (\infty,D)$ such that interaction reduces leakage, i.e., $L^A_{sum,1}(P_{X_1,Y_1,X_2},D)> L^B_{sum,2}(P_{X_1,Y_1,Z_1},D)$.
\end{thm}
\begin{IEEEproof}
Since computing the optimal $\eta^A_1$ is not straightforward, we develop lower and upper bounds for the right and left sides of (\ref{eq:eta_concavity}), respectively. Lemma \ref{lm:bound} provides a lower bound on the right-side of (\ref{eq:eta_concavity}). We now present an upper bound for the left-side of (\ref{eq:eta_concavity}). We will then compare these bounds to establish (\ref{eq:eta_concavity}).

Left-side of (\ref{eq:eta_concavity}) is given by
\begin{equation}
\begin{aligned}
\label{eq:LK3}
& \underset{P_{U|X_1}}{\text{max}}
& & H(Y_1|X_2, U) \\
& \text{subject to}
& & H(X_1|X_2,U)\leq D.
\end{aligned}
\end{equation}
Equation (\ref{eq:LK3}) is equivalent to 
\begin{equation}
\begin{aligned}
& \underset{P_{U|X}}{\text{max}}
& & H(X_1|X_2,U) + H(Y_1|X_1,X_2) - H(X_1|Y_1,X_2,U) \\
& \text{subject to}
& & H(X_1|X_2,U)\leq D.
\end{aligned}
\end{equation}
which is less than or equal to $D+ H(r)-\min_{H(X_1|X_2,U)\leq D} {H(X_1|Y_1,X_2,U)}$. We wish to show that this upper bound is strictly smaller than $C(p,q,r,\alpha_{2,0}, \alpha_{2,1})$. In fact, it suffices to show that for specific choices of $(p,q,r,\alpha_{2,0}, \alpha_{2,1)}$, a stronger bound of $C(p,q,r,\alpha_{2,0}, \alpha_{2,1}) > D+H(r)$ holds. Due to the highly parametrized nature of the source models, verifying this analytically is not straightforward. However, for specific choices of some model parameters ($p$) and a specific mechanism via parameters $\alpha_{2,0}, \alpha_{2,1}$, we can show that there exists many sources with different values of $r$ (parameter governing the public-private variable correlation) for a fixed choice of $q$ (determining the two convex choices of $X_2$). Thus, for example, for $q=0.48, p=0.7, \alpha_{2,0}=0.1, \alpha_{2,1}=0.6, r=0.23$, we can verify that $C(p,q,r,\alpha_{2,0}, \alpha_{2,1})$ is a strict lower bound on $D+H(r)=0.1853+H(0.23)$ thus establishing that there exists at least one distribution in a class of binary sources with joint distributions of the form $P_{X_1,X_2,Y_1,Y_2}$ with $Y_2=\emptyset$ for which interaction reduces leakage. Furthermore, for the same mechanism and $q$, we observe the desired non-concavity property for $p=0.79, r=0.52$. More generally, for this mechanism, as we vary $p \in (0,1)$ we can verify that there exists an $r \in (0,1)$ for which the concavity of $\eta^A_1(\cdot)$ does not hold.
\end{IEEEproof} 
 
\begin{figure}
    \begin{center}
    \includegraphics[width=10cm,keepaspectratio]{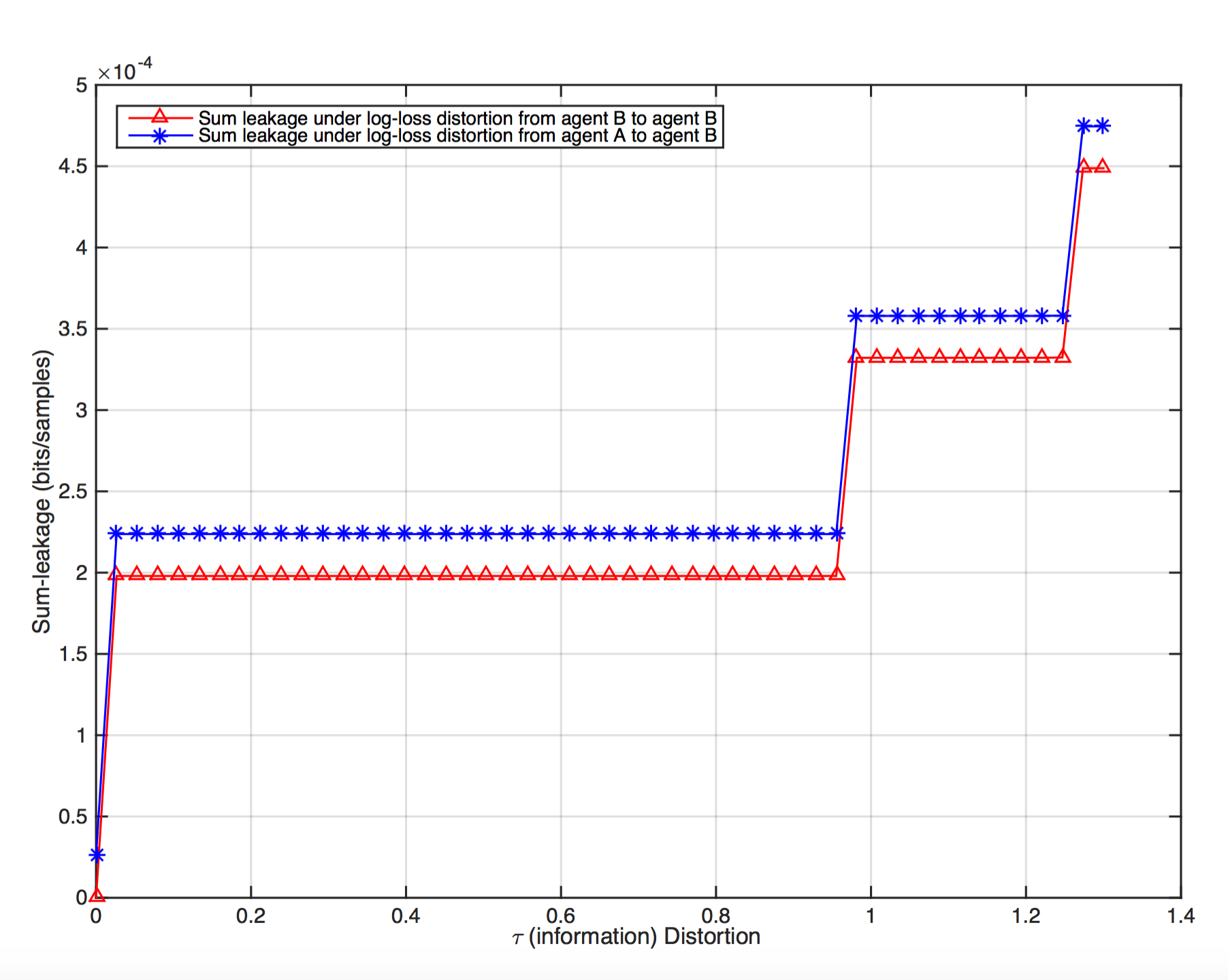}\vspace{-0.3cm}
    \caption{Comparing Sum leakage for the two round vs the one round interactive mechanism, where the blue curve with stars and the red curve with triangles show the one-round and the two-round interaction mechanism, respectively. }
    \end{center}\vspace{-0.7cm}
\end{figure} 
\section{Illustration of results}
\label{sec:sec7}

We illustrate our results for the log-loss distortion measure, and in particular, explore the effect of interaction on leakage using a publicly available dataset. The US Census dataset is a sample of US population from 1994. It contains different features including age, ethnicity, income levels, work class, and, gender such that the age feature is categorized into 7 levels, gender and income level (above 50K USD and less than 50K USD)  are binary random variables, Work class is categorized in 4 levels, and, ethnicity is classified into 4 levels. We choose $X_1 = $(age, gender), $X_2 =$ (ethnicity, gender), $Y_1 =$(work class), and, $Y_2 =$(income level), thus wishing to keep private work class and income level at agents $1$ and $2$, respectively. 

In Fig. 2, using Algorithm 1 and the empirical distribution of the data, we plot both the one-round and the two-round sum leakages as functions of mutual information based on log-loss distortion level at agent B. To demonstrate the value of interaction we consider the following results: let $d_A=0$ and $d_B$ be the log-loss distortion measure. The blue curve with stars is the leakage for one round from A to B. We note that it upper bounds the red curve with triangles which denotes the sum leakage starting from B to A and back to B, thus suggesting the interaction can reduce leakage for the log-loss setting.
\section{Conclusion}
\label{sec:sec8}
We have introduced and defined a $K$-round interactive privacy mechanism between two agents with correlated data. For the problem, we have determined the leakage-distortion region for general distortion functions, and in particular focused on the log-loss distortion measure. For specific distortion measures and source distributions as well as for sources under log-loss distortion, we have illustrated that interaction can reduce leakage. This model captures multiple applications from interaction in critical infrastructure to distributed deep learning in which data has to be shared between multiple computing systems while ensuring privacy of appropriate features. To this end, our composition result highlights how to design mechanisms for each rounds that are aware of the sharing in the previous rounds to ensure no additional leakage and even a reduction in leakage relative to a one-shot setting. Future work will include extending to multiple agents, data-driven empirical and approximate approaches (e.g., canonical mechanisms that can model many different source classes), as well as  evaluating leakage for different classes of statistical inference attacks.

\appendices
\section{Proof of Theorem \ref{m:main}}
We use the following notion of strong typicality introduced in \cite[p. 25]{ElGamal2011}. For a random variable $X \sim p(x)$ and $\epsilon \in (0,1)$, the set $\mathcal{T}_{\epsilon}^{(n)}(X)$ of $\epsilon$-typical sequences is
\begin{flalign}
\mathcal{T}_{\epsilon}^{(n)}(X)=\{x^n: |\pi(x|x^n) - p(x)| \le \epsilon{p(x)}, \forall x \in \mathcal{X}\} \label{eq:e_typical}
\end{flalign}
where the empirical probability mass function of $x^n$, i.e., type, for every $x \in \mathcal{X}$ is defined as
\begin{flalign}
\pi(x|x^n) = \frac{|\{i:x_i=x\}|}{n}.
\end{flalign}
{We also use the corresponding notations for jointly and conditionally typical sequences as in \cite[p. 27]{ElGamal2011}. Thus, the definition in (\ref{eq:e_typical}) for the set of typical sequences of a single random variable can be generalized to the set of jointly typical sequences of $(X^n,Y^n)$, i.e., $\mathcal{T}_{\epsilon}^{(n)}(X,Y)$, by viewing $(X,Y)$ as a single ``large" random variable \cite[p. 27]{ElGamal2011}. Then, the set of conditionally $\epsilon$-typical $n$-length sequences for a random variable $X$ conditioned on $Y^n=y^n$ is $\mathcal{T}_{\epsilon}^{(n)}(X|y^n)=\{x^n:(x^n,y^n) \in \mathcal{T}_{\epsilon}^{(n)}(X,Y)\}$.}

\begin{IEEEproof}
\underline{Achievability:} For ease of exposition, let $\mathbf{W} = (U^n_1,\dots, U^n_K, X^n_2)$ with the $i^{th}$ entry $W_i = (U_{1,i},\dots, U_{K,i}, X_{2,i})$. {Our proof below relies on the fact that we consider a sequence of $K$ memoryless mechanisms as an achievable scheme}. The resulting leakage over $K$-rounds is
\begin{flalign}
&I(Y^n_1; U^n_1,\dots,U^n_K,X^n_2)\notag\\
&=nH(Y_1)- H(Y^n_1|\mathbf{W})\label{eq:iid}\\
&=nH(Y_1) -\sum_{\mathbf{w}}{P(\mathbf{w})}H(Y^n_{1}| \mathbf{W}=\mathbf{w})\label{eq:def2}\\
&=nH(Y_1)-\sum_{\mathbf{w}\in\mathcal{T}_{\epsilon}^{(n)}(\textbf{W})}{P(\mathbf{w})}{{H(Y^n_{1}| \mathbf{w})}}-\sum_{\mathbf{w}\not\in\mathcal{T}_{\epsilon}^{(n)}(\textbf{W})}{P(\mathbf{w})}{{H(Y^n_{1}| \mathbf{w})}}\label{eq:def3}\\
&\le nH(Y_1)-\sum_{\mathbf{w}\in\mathcal{T}_{\epsilon}^{(n)}(\textbf{W})}{P(\mathbf{w})}{{H(Y^n_{1}| \mathbf{w})}} \label{eq:def3a}\\
&\le nH(Y_1) - \sum_{\mathbf{w}\in\mathcal{T}_{\epsilon}^n({\mathbf{W}})}{P(\mathbf{w})}\sum_{\mathbf{y}_{1}\in\mathcal{T}_{\epsilon}^n({{Y_{1}}|\mathbf{w}})}P(\mathbf{y}_{1}|\mathbf{w}){\log(2^{-n(H(Y_1|W)+\epsilon(n))})}\label{eq:def4b}\\
&-\sum_{\mathbf{w}\in\mathcal{T}_{\epsilon}^n(\mathbf{W})}{P(\mathbf{w})}\sum_{\mathbf{y}_{1}\not\in\mathcal{T}_{\epsilon}^n({{Y_{1}}|\mathbf{w}})}-P(\mathbf{y}_{1}|\mathbf{w})\log(P(\mathbf{y}_{1}|\mathbf{w})) \notag\\
&= nH(Y_1)-nH(Y_1|W) - \epsilon(n)\label{eq:def5}\\
&-\sum_{\mathbf{w}\in\mathcal{T}_{\epsilon}^n(\mathbf{W})}{P(\mathbf{w})}\sum_{\mathbf{y}_{1}\not\in\mathcal{T}_{{Y_{1}}|\mathbf{w}}}-P(\mathbf{y}_{1}|\mathbf{w})\log(P(\mathbf{y}_{1}|\mathbf{w}))\notag\\
&\le nI(Y_1;U_1,\dots,U_K,X_2) - 2\epsilon(n)\label{eq:def6}
\end{flalign}
{where $\mathcal{T}_{W}=\mathcal{T}_{U_1,\dots,U_K,X_2}$ and $\mathcal{T}_{{Y_1}|w}=\mathcal{T}_{{Y_{1}}|u_1,\dots,u_K,x_2}$ are sets of jointly typical sequences and conditional typical sequences, respectively, (\ref{eq:def3a}) follows from the fact that the entropy, conditional or otherwise, of a discrete random variable is non-negative, (\ref{eq:def4b}) follows from using a memoryless mechanism and lower bounding the probability of a typical $Y^n$ sequence conditioned on a typical $\mathbf{w}$ {as $2^{-n(H(Y_1|W)+\epsilon(n))}$}, and finally (\ref{eq:def6}) follows from the fact that the non-typical set has asymptotically vanishing probability for sufficiently large $n$ (implicit in (\ref{eq:def6}) is the fact that since the non-typical set is a measure zero set asymptotically, its entropy also approaches zero). Note that both $\epsilon(n)$ and $\delta(n)$ go to zero as $n$ goes to infinity.}

Finally, it can be verified in a straightforward manner that the distortion for memoryless mechanisms simplifies to the single-letter expression in (\ref{achievable requirement3}).

\underline{Converse:}
Let $D_{j,i}$, $j=1,2$, $i=1,2,\dots,n$, denote the distortion of the $i^{th}$ data entry of source $X^n_j$. Given a $K$-round mechanism with outputs $U^n_1,\dots,U^n_K$, we have 
\begin{flalign}
\label{eq:e1}
L_1+\epsilon&\geq\frac{1}{n}I(Y^n_1;U^n_1\dots,U^n_K,X^n_2)\\
\label{eq:e2}
&= \frac{1}{n}[I(Y^n_1; X^n_2)+I(Y^n_1;U^n_1\dots,U^n_K|X^n_2)]\\
&=\frac{1}{n}\sum_{i=1}^{n}[I(Y_{1i}; X_{2i}) \label{eq:e3}\\
&+H(Y_{1i}|X_{2i}) - H(Y_{1i}|U^n_1\dots, U^n_K,X^n_2, Y^{i-1}_1)]\notag\\
&\geq\frac{1}{n}\sum_{i=1}^{n}[I(Y_{1i}; X_{2i})\label{eq:e4}\\
&+H(Y_{1i}|X_{2i}) - H(Y_{1i}|U_{1i}\dots, U_{Ki},X_{2i})] \notag\\
& = \frac{1}{n}\sum_{i=1}^{n}I(Y_{1i};U_{1i}\dots,U_{Ki},X_{2i})\\
& = \frac{1}{n}\sum_{i=1}^{n}L_{U,1}(D_{1i},D_{2i}) \label{eq:e5}\\
& \ge L_{U,1}(D_{1},D_{2}) \label{eq:e6}
\end{flalign}
where (\ref{eq:e1}) follows from (\ref{achievable requirement3}), (\ref{eq:e2}) uses chain rules for mutual information, (\ref{eq:e3}) follows from chain rule and the fact that sources are i.i.d., (\ref{eq:e4}) follows from the fact that condition reduces entropy, (\ref{eq:e5}) follows from the definition of $L_{U,1}$ in (\ref{eq:LU_def}) {and from identifying $D_{1i}$ and $D_{2i}$ as the distortion for the $i^{th}$ entries of $X_1^n$ and $X_2^n$, respectively,} and finally (\ref{eq:e6}) follows from the fact that leakage-distortion function can be shown to be a non-increasing and convex function of the distortion pair $(D_1,D_2)$\cite{Yamamoto1983,Sankar2013}.

From the definition of the mechanism, $Y^n_{j}\leftrightarrow (U^n_{1},\dots, U^n_{2k-1}, X^n_{k})\leftrightarrow U^n_{2k}$, $j=1,2, k=1,2, j \ne k,$ forms a Markov chain. Then, for memoryless sources, the following Markov chains hold:
\begin{flalign}
\label{eq:m1}
Y_{1i}\leftrightarrow (U_{1i},\dots, U_{2k-1,i}, X_{2i})\leftrightarrow U_{2k,i}\\
\label{eq:m2}
Y_{2i}\leftrightarrow (U_{1i},\dots, U_{2k-2,i}, X_{1i})\leftrightarrow U_{2k-1,i}.\vspace{-0.4cm}
\end{flalign}
{Finally, the cardinality bounds for $\mathcal{U}_1,...,\mathcal{U}_K$ can be derived using standard methods that rely on Caratheodory's theorem and are omitted for reasons of space.}

\end{IEEEproof}

\section{Proof of (\ref{eq:ErasureLeak})}
 \begin{IEEEproof}
  Setting $K=1$ in (\ref{eq:ksum}) and using the fact that $Y_2 \leftrightarrow X_1 \leftrightarrow U_1$ forms a Markov chain from (\ref{eq:MC_1}), we have $L^A_{sum,1}(D_1,0)=\min_{P_{U_1|X_1}} [I(X_1;Y_2)+I(Y_1;U_1,X_2)]$. For the given sources, this simplifies as
\begin{flalign}
\label{eq:AppB}
&L^A_{sum,1}(D_1,0)=2- H(2p(1-p))-\max_{{P_{U_1|X_1}}}{H(Y_1|U_1, X_2)}.
\end{flalign}

For finite average distortion under erasure distortion in (\ref{eq:erasuredist}), it suffices to consider $P_{U_1|X_1}$ with
\begin{equation}
P_{U_1|X_1} = \begin{cases} 
      \alpha_0, & \textrm{ if $x=0$ and $u=e$} \\
      1-\alpha_0, & \textrm{ if $x=0$ and $u=0$} \\
      \alpha_1, & \textrm{  if $x=1$ and $u=e$}\\
      1-\alpha_1, & \textrm{ if $x=1$ and $u=1$} \\
      0, & \textrm{ otherwise} \\
   \end{cases} 
   \end{equation}
where $\mathbb{E}(d_1(X_1,U_1))=P_{X_1}(0)\alpha_0+P_{X_1}(1)\alpha_1\le D_1$. 
For the reconstruction function, note that since the output alphabet allows erasure, for finite average distortion we require $\hat{X}_1(u=e,x_2 = 0) = \hat{X}_1(u=e,x_2 = 1) = e$; for the same reason, we require that if $u=0,1$, then $\hat{X}_1(u,x_2=i)$ takes values either $e$ or $i$ but not $1-i$. Of all such mappings, one can verify that a mapping $\hat{X}_1=U$ achieves the minimal distortion for the distortion function in (\ref{eq:erasuredist}). Since $\hat{X}_1$ is a function of $U_1,X_2$, we have that $H(Y_1|U_1,X_2)=H(Y_1|U_1,X_2,\hat{X}_1) \le H(Y_1|X_2,\hat{X}_1=U_1)$. For this choice of $\hat{X}_1=U_1$, we can write the achievable distortion as a function of $P_{U_1|X_1}$. On the other hand, for the leakage function, we use the joint distribution of $(X_1,X_2,Y_1,Y_2)$ to expand $H(Y_1|U_1, X_2)$ as a function of $P_{Y_1|U_1,X_2}$ as
\small
\begin{flalign}
H(Y_1|U_1, X_2) = &\frac{1}{2} (1-\alpha_0)H(p)+\frac{1}{2} (1-\alpha_1)H(p)\\
+&[\frac{\alpha_0}{2}{(1-p)}+\frac{\alpha_1}{2}{p}]H(\frac{(1-p)^2\alpha_0+p^2\alpha_1}{(1-p)\alpha_0+p\alpha_1})\\
+&[\frac{\alpha_0}{2}{p}+\frac{\alpha_1}{2}{(1-p)}]H(\frac{p(1-p)\alpha_0+p(1-p)\alpha_1}{p\alpha_0+(1-p)\alpha_1}).
\end{flalign}
\normalsize
We observe that, for a fixed source distribution, $H(Y_1|U_1,X_2)$ is concave with respect to $P_{X_1,Y_1,X_2,Y_2,U_1}$ and also $P_{U_1|X_1}$. Furthermore, since $P_{U_1|X_1}$ is linear with respect to $\alpha_{0},\alpha_{1}$, $H(Y_1|U_1,X_2)$ is also concave with respect to $P_{U_1|X_1}$. Upon simplification of both the average distortion using (\ref{eq:erasuredist}) and the objective function in (\ref{eq:AppB}), we find that both functions are symmetric with respect to  $(\alpha_{0},\alpha_{1})$; in particular, we have that $H(Y_1|U_1, X_2)$ is maximized if $\alpha_0 = \alpha_1 = \alpha$, which in turn leads us to our result.

\end{IEEEproof}






%
\bibliographystyle{IEEEtran}
\bibliography{InteractivePrivacy_Refs}
%








\end{document}